\documentclass[reprint,longbibliography,aip]{revtex4-1}

\usepackage{graphicx} 
\usepackage{amsmath}
\usepackage{ amssymb }

\usepackage{color}

\usepackage{float} 
\usepackage{wrapfig} 

\usepackage[makeroom]{cancel} 
\usepackage{comment}

\usepackage{lipsum} 

\usepackage{outlines}

\newcommand{\beq}{\begin{equation}}
\newcommand{\eeq}{\end{equation}}
\newcommand{\vv}{\mathbf{v}}

\newcommand{\bvec}{\begin{pmatrix}}
\newcommand{\evec}{\end{pmatrix}}
\newcommand{\lp}{\left(}
\newcommand{\rp}{\right)}
\newcommand{\pa}[2]{\frac{\partial #1}{\partial #2}}

\newcommand{\paf}[2]{\partial #1 / \partial #2}
\newcommand{\llangle}{\left \langle}
\newcommand{\rrangle}{\right \rangle}

\newcommand{\pv}{\text{PV}}

\newcommand{\ve}[1]{\mathbf{#1}}

\newcommand{\tn}{\tilde{n}}
\newcommand{\tv}{\tilde{v}}

\newcommand{\tphi}{\tilde{\phi}}
\newcommand{\tE}{\tilde{E}}
\newcommand{\sgn}{\text{sgn}}

\newcommand{\tx}{\tilde{x}}
\newcommand{\dew}[1]{\boldsymbol{\mathcal{#1}}}

\renewcommand{\Re}[1]{\text{Re} #1}

\newcommand{\tu}{\tilde{u}}

\newcommand{\gammab}{\bar{\gamma}}
\newcommand{\etab}{\bar{\eta}}

\newcommand{\bx}{\bar{x}}

\newcommand{\bp}{\bar{p}}
\newcommand{\bH}{\bar{H}}
\newcommand{\bphi}{\bar{\phi}}

\AtBeginDocument{%
	\newwrite\bibnotes
	\def\bibnotesext{Notes.bib}
	\immediate\openout\bibnotes=\jobname\bibnotesext
	\immediate\write\bibnotes{@CONTROL{REVTEX41Control}}
	\immediate\write\bibnotes{@CONTROL{%
			aip41Control,author="08",editor="1",pages="1",title="0",year="1"}}
	\if@filesw
	\immediate\write\@auxout{\string\citation{aip41Control}}%
	\fi
}%

\begin{document}



\title{Momentum Conservation in Current Drive and  Alpha-Channeling-Mediated Rotation Drive}

\author{Ian E. Ochs}
\email{iochs@princeton.edu}
\affiliation{Department of Astrophysical Sciences, Princeton University, Princeton, New Jersey 08540, USA}
\author{Nathaniel J. Fisch}
\affiliation{Department of Astrophysical Sciences, Princeton University, Princeton, New Jersey 08540, USA}

\date{\today}

\begin{abstract}

Alpha channeling uses waves to extract hot ash from a fusion plasma, while transferring energy from the ash to the wave. Intriguingly, it has been proposed that the extraction of this charged ash could create a radial electric field, efficiently driving $\ve{E} \times \ve{B}$ rotation. However, existing theories ignore the response of the nonresonant particles, which play a critical role in enforcing momentum conservation in quasilinear theory. Because cross-field charge transport and momentum conservation are fundamentally linked, this non-consistency throws the whole effect into question.

Here, we review recent developments that have largely resolved this question of rotation drive by alpha channeling. We build a simple, general, self-consistent quasilinear theory for electrostatic waves, applicable to classic examples such as the bump-on-tail instability. As an immediate consequence, we show how waves can drive currents in the absence of momentum injection even in a collisionless plasma. To apply this theory to the problem of ash extraction and rotation drive, we develop the first linear theory able to capture the alpha channeling process. The resulting momentum-conserving linear-quasilinear theory reveals a fundamental difference between the reaction of nonresonant particles to plane waves that grow in time, versus steady-state waves that have nonuniform spatial structure, allowing rotation drive in the latter case while precluding it in the former. This difference can be understood through two conservation laws, which demonstrate the local and global momentum conservation of the theory. Finally, we show how the oscillation-center theories often obscure the time-dependent nonresonant recoil, but ultimately lead to similar results.

\end{abstract}

\maketitle

\section{Background and Introduction}

A deuterium-tritium fusion reaction produces fast neutrons and hot helium ash.
While the neutrons quickly leave the device, the charged ash remains confined in the reactor for a time.
Initially, this confinement is somewhat beneficial, as the ash is extremely hot ($3.5$ MeV) compared to the bulk plasma ($\sim 20$ keV), and thus transfers heat to the bulk via collisions.

However, this situation is not ideal for a few reasons.
First, the radial gradient of hot alpha particles represents a large source of free energy, which can drive plasma instabilities.
Second, the hot ash primarily heats electrons.
Although some of the heat eventually makes its way to ions via electron-ion collisions, this energy flow pattern sets up a situation where the electrons are hotter than the ions, which increases pressure and radiative energy loss without increasing the fusion power.
Third, once the ash thermalizes, it no longer provides any heat, but quasineutrality requires that each alpha particle be accompanied by two electrons, further increasing the pressure without additional fusion power.

For all of these reasons, it is desirable to quickly extract the alpha particles from the plasma, while redirecting their energy into useful work, such as ion heating and current drive. 
Since both of these tasks can be accomplished by suitably chosen plasma waves, it makes sense to try to transfer the ash energy into a these waves.

\begin{figure*}
	\center
	\includegraphics[width=0.9\linewidth]{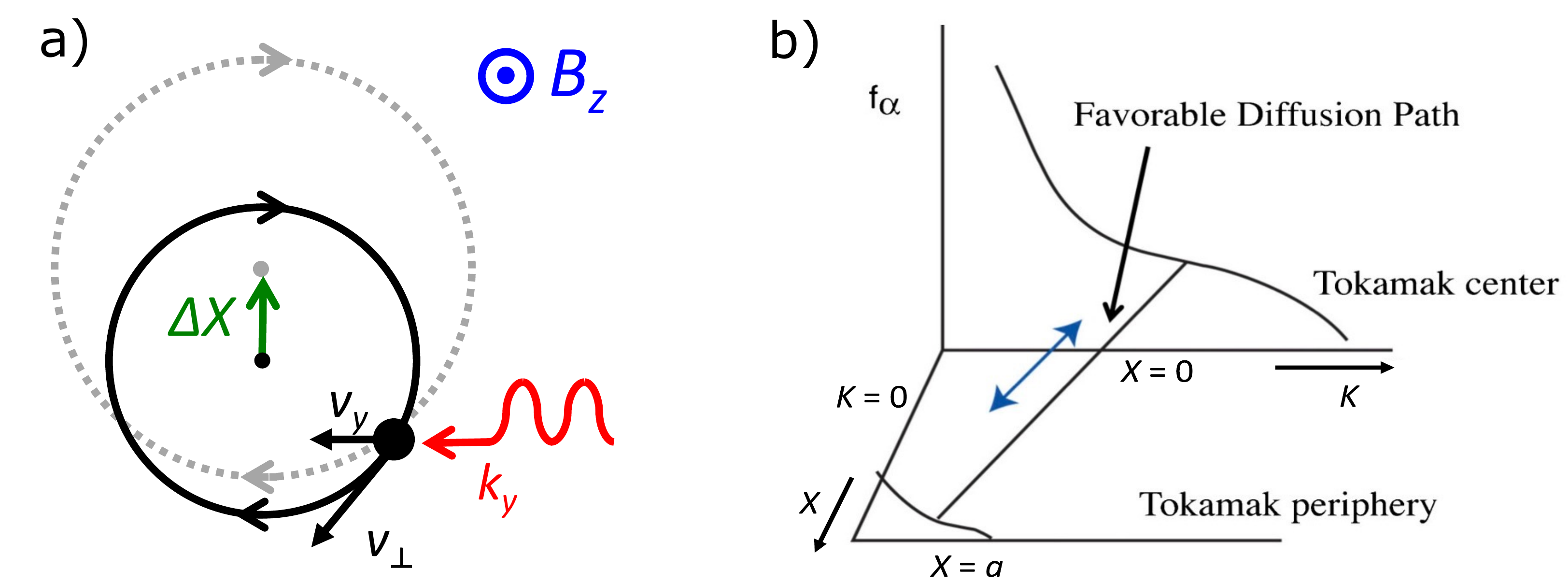}
	\caption{
		Schematic view of alpha channeling by a lower hybrid wave.
		(a) As a hot alpha particle orbits in a magnetic field $B_z$, it becomes Landau resonant with the LH wave twice per orbit. At this resonance point, it receives a ``kick,'' either gaining or losing energy. Correlated with this change in energy, the center of its orbit moves as well, resulting in a coupled random walk (diffusion) along a 1D path in energy ($K$) and space ($X$).
		(b) By choosing a wave where this diffusion path connects high-energy particles at the tokamak center (an alpha particle source) to low-energy particles at the edge (a sink), alpha ash can be made to diffuse out while transferring its energy into the wave. Figure (b) is adapted from Ref.~\cite{Fisch2014}.
	}
	\label{fig:alphaChannelingSchemIntro}
\end{figure*}

Such a flow of energy from the alpha particle ash into the waves, as in the case of instabilities, is made possible by the free energy associated with the radial gradient of the alpha particles, which are born at the hot reactor core.
A suitably chosen wave can tap this free energy by pulling ash outward along this gradient, both capturing the ash energy and removing the cold ash from the reactor, accomplishing two goals with a single process. 
Because this generic process involves channeling energy from alpha particles into waves, while also channeling the alpha particles out of the reactor, it has been termed ``alpha channeling.''

A blueprint for alpha channeling was first proposed by Fisch and Rax \cite{Fisch1992,fisch1992current}, using the electrostatic lower hybrid (LH) wave.
Sufficiently strong LH waves are useful for both current drive and fast ion heating, and also interact strongly with fusion-born alpha particles.
While we will focus here on alpha channeling via lower hybrid waves \cite{Fisch1992,fisch1992current,Heikkinen1996,ochs2015coupling,ochs2015alpha}, it should be noted that one can also make use of waves in the ion-cyclotron range of frequencies \cite{Valeo1994,fisch1995ibw,Fisch1995a,Heikkinen1995,Marchenko1998,Kuley2011,Sasaki2011,Chen2016a,Gorelenkov2016,Cook2017,Castaldo2019,Romanelli2020}, and can further optimize the effect by combining multiple waves \cite{Herrmann1997,Cianfrani2018,Cianfrani2019,White2021}.

Alpha particles interact with LH waves via a Landau resonance, wherein twice per orbit in a magnetic field, the velocity of the alpha particle matches the phase velocity of the wave (Fig \ref{fig:alphaChannelingSchemIntro}a).
Karney \cite{Karney1978,karney1979stochastic} showed that for sufficiently strong LH waves, this process leads to a diffusion in energy, with particles randomly gaining or losing energy with each interaction with the wave.
Taken alone, i.e. in a homogeneous plasma, this diffusion leads to net transfer of energy from the wave into the particles, which is undesirable for alpha channeling.

What Fisch and Rax showed was that this diffusion in energy is coupled to diffusion in the position of the particle, so that particles which gain energy move in one direction, while particles that lose energy move in the opposite direction (Fig.~\ref{fig:alphaChannelingSchemIntro}a).
Formally, the overall diffusion thus takes place on a one-dimensional path in energy-position space.
Thus, by creating a path that connects high-energy particles at the fusing core of the plasma (a source) with low-energy particles at the plasma edge (a sink), the diffusion can be made to extract alpha particles while cooling them (Fig.~\ref{fig:alphaChannelingSchemIntro}b), thus accomplishing the goal of alpha channeling.

During this process of wave-mediated ash extraction, it is natural to hypothesize that the \emph{charge} of the ash is pulled out as well.
Such charge extraction would create a radial electric field in the magnetized plasma, resulting in $\ve{E} \times \ve{B}$ rotation.
Driving $\ve{E} \times \ve{B}$ rotation, and in particular sheared rotation, is useful in a variety of ways for plasma control; it can suppress both large-scale instabilities \cite{Shumlak1995,Shumlak2003,golingo2005deflagration,Zhang2019,Huang2001,Ellis2001,Ellis2005,Ghosh2006} and small-scale turbulence \cite{Taylor1989,maggs2007transition,Carter2009,Burrell2020,DiSiena2021}, increase the confinement of mirror plasmas via centrifugal forces\cite{Lehnert1971,bekhtenev1980problems,Teodorescu2010}, and even be exploited in a variety of plasma mass separation schemes \cite{gueroult2014plasma,gueroult2015plasma, Gueroult2018ii,Zweben2018,ohkawa2002band,litvak2003archimedes,gueroult2014double,hellsten1977balance,ochs2017drift,fetterman2011magnetic}.
Thus, rotation drive via alpha channeling would be an extremely useful tool to have in the plasma control toolbox.

Such rotation due to alpha channeling was first proposed by Fetterman and Fisch \cite{fetterman2008alpha,fetterman2010stationary,fetterman2012wave}.
In particular, they proposed that for a rotating plasma, a stationary electrostatic or magnetostatic wave at the plasma edge would appear as a propagating wave in the rotating plasma reference frame, pulling out alpha particles and putting their energy into the plasma rotation\cite{fetterman2010stationary}.
This would result in a very efficient direct conversion of energy from fusion ash heat into plasma rotation.

While this picture is very appealing, the charge extraction proposal of Fetterman and Fisch must be looked at as a hypothesis, since the theory that they built only looked at the response of the Landau-resonant alpha particles to the wave, without considering the response of the bulk plasma.
Thus, embedded in the theory is an assumption: the bulk particles will not move across magnetic field lines in response to the wave, leaving only the alpha particle charge transport.

There is a puzzle here, however.
Consider the case of an electrostatic plane wave.
Such a wave has no electromagnetic momentum, which is given by the Poynting flux, $\ve{p} = \ve{S} / c^2 = \ve{E} \times \ve{B} / 4\pi c$.
Thus, as the wave amplifies due to alpha channeling, momentum must be conserved in the plasma.

\begin{figure}[t]
	\center
	\includegraphics[width=0.7\linewidth]{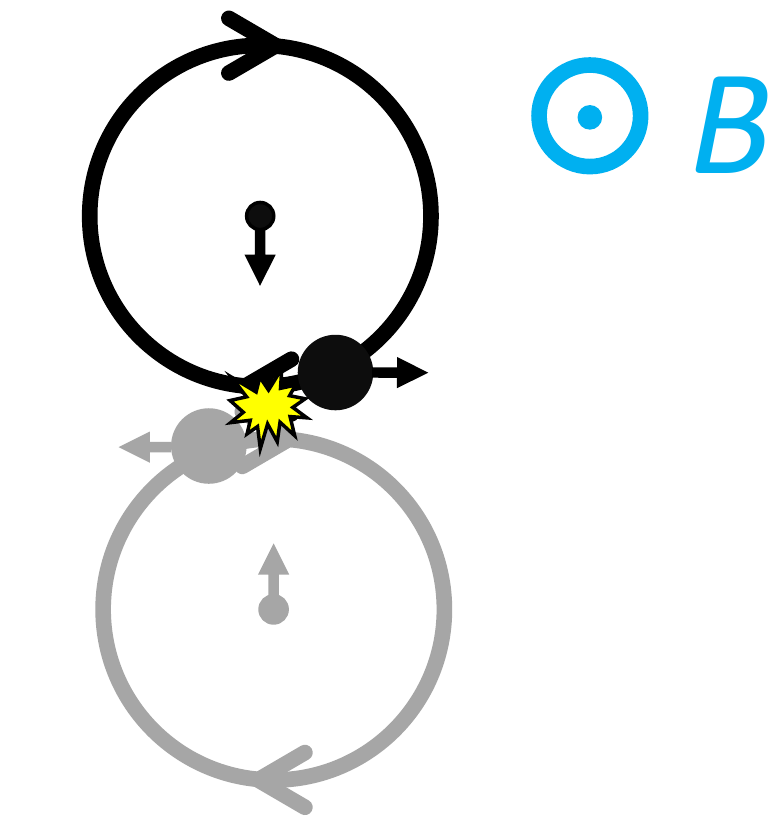}
	\caption{
		When two particles collide in a uniform magnetic field, their gyrocenters move such that no net charge moves.
		This cancellation ultimately comes from the fact that momentum is conserved in the collision.
		Because a planar electrostatic wave has no electromagnetic momentum, as the wave amplifies or damps, momentum should be conserved in the plasma.
		Thus, the wave amplification or damping can be seen as mediating a momentum-conserving ``collision'' between particles in the plasma, making charge transport questionable.
	}
	\label{fig:momentumCollision}
\end{figure}

This momentum conservation has implications for the transport of charge, which is most familiar from the theory of classical collisional transport in magnetized plasmas.
When two particle collide in a uniform magnetic field, their gyrocenters move in just such a way that no net charge moves across field lines (Fig.~\ref{fig:momentumCollision}).
As we discuss in Section~\ref{sec:momentumMagnetized}, this cancellation of net charge transport is intrinsically linked to the fact that the collision conserves momentum.
This link suggests that we must tread with care in arguing that alpha channeling pulls charge across field lines, and thus drives rotation.

What is ultimately needed is a self-consistent theory of alpha channeling, treating both resonant and nonresonant particles.
Such a theory has been recently developed \cite{Ochs2020,ochs2021nonresonant,Ochs2021WaveDriven,Ochs2022Thesis} for the simple case of alpha channeling in a slab-geometry plasma, and the main purpose of this paper is to concisely review these developments.
However, this paper will also discuss the comparison of this new theory to other theories of ponderomotive forces, discussing the advantages and disadvantages of each approach.

We begin in Section~\ref{sec:momentumMagnetized} by reviewing the deep link between momentum conservation and charge transport in a magnetized plasma \cite{HelanderSigmar,kolmes2019radial}. 
To eliminate confounding processes that might muddy the subsequent analysis, we work in the simplest possible magnetic geometry: a slab with a uniform magnetic field.
In such a geometry, we show that large radial charge transport requires an azimuthal force to be applied to the plasma.
In Section~\ref{sec:conservation1D}, we review the momentum conservation principles in plasma wave problems, showing how planar electrostatic waves cannot produce a net force on the plasma.
In Section~\ref{sec:btiQL}, we see how these conservation laws play out in the quasilinear theory of the bump-on-tail instability.
This familiar textbook \cite{kralltrivelpiece,davidson1973methods,Tsytovich1977} example shows the importance of the nonresonant response in enforcing momentum conservation \cite{Kaufman1972reformulation}.
Analysis of this case allows us to derive a simple form of the quasilinear force, consistent with momentum conservation and the kinetic theory of the bump-on-tail instability, but valid for any planar electrostatic wave in a homogeneous plasma.

As an immediate application, this force reveals how currents can be driven by ion acoustic waves in plasmas despite the nonresonant recoil, even in the absence of collisions.
We explore this current drive in Section~\ref{sec:currentDrive}, which can be skipped by readers interested only in alpha channeling and rotation drive.

With our groundwork laid, we return in Section~\ref{sec:alphaIvp} to the problem of alpha channeling.
We discuss what has historically made a self-consistent linear and quasilinear theory of alpha channeling challenging, and how to get around these issues.
By applying our simple electrostatic theory of the quasilinear force, we show how in the plane-wave initial value problem of alpha channeling, the resonant charge extraction identified by Fetterman \cite{fetterman2008alpha} is canceled out by a nonresonant charge transport in the opposite direction, eliminating the rotation drive effect.

In Section~\ref{sec:alphaBvp}, we turn our attention to the problem in multiple dimensions.
We extend our quasilinear theory to this case, discussing how the addition of spatial fluxes to the conservation laws changes the fundamental constraints of the problem.
In particular, we show how the Reynolds and Maxwell stress combine to cancel the nonresonant charge transport, allowing resonant charge extraction and rotation drive consistent with local momentum conservation.
Furthermore, using a Fresnel analysis and our conservation laws, we show how the force on the plasma needed to extract these particles is ultimately provided by the antenna that launches the wave.

Finally, in Section~\ref{sec:ocComp}, we compare our analysis to the oscillation center theories of ponderomotive forces \cite{Dodin2014Variational,Dewar1973}.
These theories can make the recoil force difficult to calculate, but also offer some general perspectives on the steady-state problem.

\section{Momentum Conservation in a Magnetized Plasma Slab} \label{sec:momentumMagnetized}

To examine perpendicular charge transport, we consider the simplest possible plasma setup.
Namely, we consider a slab plasma, uniform along $y$ and $z$, with a magnetic field $\ve{B}$ along the axis $\hat{z}$. 
Any quantities which vary in space (such as the electric potential) vary only along the $x$ direction.
Thus, $x$ can be thought of as the ``radial'' coordinate, $y$ as the ``poloidal'' coordinate, and $z$ as the ``axial'' or ``toroidal'' coordinate.
We are interested in charge transport along $x$, which will lead to a change in the $y$-directed $\ve{E} \times \ve{B}$ drift, i.e. the plasma rotation.

Consider the motion of a charged particle in such a uniform magnetic field.
The magnetic field is given by the vector potential $\ve{A} = B x \hat{y}$.
Thus, ignoring the $\hat{z}$ direction and working in the $x$-$y$ plane, the particle motion exhibits two invariants, from the energy and $y$-directed canonical momentum:
\begin{align}
	H &= \frac{1}{2} m v^2 + \phi(x)\\
	p_y &= m v_y + \frac{q B}{c} x.
\end{align}
The upper and lower bounds of the orbit can be used to find the gyrocenter.
When $\phi(x)$ is a linear function of $x$, i.e. when the radial electric field is constant, then the gyrocenter position is simply given by:
\begin{align}
	x_\text{gc} &= \frac{c}{q B} p_y.
\end{align}

Now consider a collision between any number of particles, but which conserved momentum.
The total cross field charge transport will be given by:
\begin{align}
	\sum_i q_i \Delta x_{\text{gc},i} &= \frac{c}{B} \sum_i \Delta p_{y,i} = 0.
\end{align}
Thus, there is no net movement of charge due to the collision: movement of net charge requires momentum input, i.e. a net force on the plasma.

\subsection{A Note on Viscosity}
Interestingly, this conclusion relaxes if the electric field varies in $x$.
As we show in Appendix~\ref{app:viscosity}, the relation between gyrocenter and energy becomes nonlinear as soon at the electric field has a nonzero second derivative, allowing for net charge transport in a momentum-conserving collision.
However, this charge transport is ordered down by $\epsilon^2$ relative to the motion of each individual charge, where $\epsilon \sim \rho_i / L$, with $\rho_i$ the gyroradius and $L$ the characteristic scale length of the electric field variation.
This scaling leads to cross field-currents consistent with the Braginskii perpendicular viscosity \cite{Braginskii1965,kolmes2019radial} and collisional gyrokinetics \cite{ParraCatto2008CollisionalGyrokinetics,Krommes2013}, and can be seen as a Larmor-radius-scale random walk of momentum.

A broader discussion of the multiple ways to view these small cross-field currents, including in the two-fluid \cite{kolmes2019radial}, MHD, and particle pictures, is given in Ref.~\cite{Ochs2022Thesis}.
However, for the purposes of this paper, we will be focused on whether alpha channeling leads to \emph{uncompensated} charge transport, i.e. to 0th order net $x$-directed currents independent of the flow shear variation length $L$.
The present analysis suggests that this requires a net force to be applied to the plasma along the $y$ direction: a problem for a planar electrostatic wave, which has no momentum.

\section{Conservation Laws}\label{sec:conservation1D}

Before launching into a discussion of momentum conservation in waves problems, it is good to clarify what is meant by momentum (and, for that matter, energy).
In plasma waves, two types of energy and momentum are common to discuss.
The first are the energy and momentum associated with the electromagnetic fields in the plasma:
\begin{align}
	W_{EM} = \frac{E^2 + B^2}{8\pi}; \quad p_{EM}^i = \frac{1}{4\pi c}\epsilon^{ijk} E_j B_k \label{eq:wpEM}.
\end{align}
These form a closed system with \emph{all} the particles constituting the plasma, which (in the sub-relativistic limit) have energy and momentum:
\begin{align}
	W_{P} = \int  \frac{1}{2} m v^2  f d \ve{v}; \quad p_{P}^i = \int  m v^{i} f  d \ve{v} \label{eq:wpP},
\end{align}
where $f(t,\ve{x},\ve{v})$ is the distribution function.
In a homogeneous plasma, this conservation manifests as:
\begin{align}
	\frac{d}{dt} \lp W_{EM} + W_P \rp =0; \quad \frac{d}{dt} \lp p^i_{EM} + p^i_P \rp =0. \label{eq:cons1DEMP}
\end{align}

The second form of energy and momentum often discussed is known as the generalized Minkowski \cite{dodin2012axiomatic} or plasmon \cite{Tsytovich1977} energy and momentum, which incorporates the oscillating motion of the plasma particles:
\begin{align}
	W_M = \omega_r \mathcal{I}; \quad p^i_M = k_r^i \mathcal{I} \label{eq:wpMinkowski}.
\end{align}
Here, $\omega_r$ and $k_r$ are the real components of the wave frequency and wavenumber, and $\mathcal{I}$ is the wave action, given for electrostatic waves by:
\begin{align}
	\mathcal{I} &= W_{EM} \pa{D_r}{\omega_r}, \label{eq:actionES}
\end{align}
where $D_r$ is the real part of the dispersion function, defined in detail in Section~\ref{sec:btiQL}.
Although the Minkowski energy and momentum do not generally form a part of a closed system, in a collisionless, homogeneous plasma, they form a closed system with the \emph{resonant} particles, in the sense that:
\begin{align}
	\frac{d}{dt} \lp W_{M} + W_{RP} \rp =0; \quad \frac{d}{dt} \lp p^i_{M} + p^i_{RP} \rp =0. \label{eq:cons1DMRP}
\end{align} 
The resonant particle energy and momentum are given in the same way as the nonresonant energy and momentum from Eqs.~(\ref{eq:wpP}), but with the integration only over the resonant region.

\begin{figure}[t]
	\center
	\includegraphics[width=0.9\linewidth]{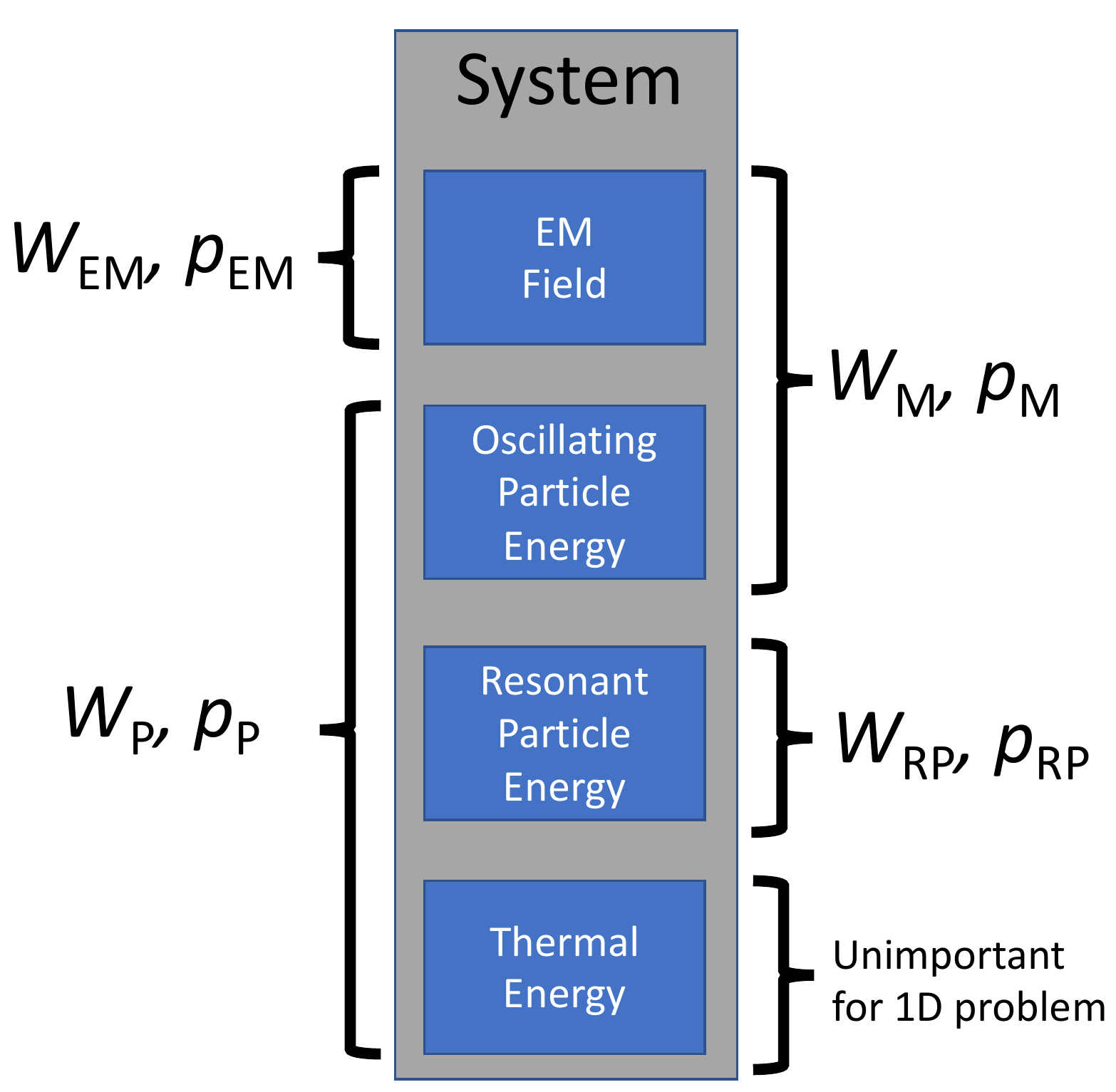}
	\caption{
		Schematic depiction of the difference between electromagnetic and Minkowski energy and momentum.
		The Minkowski energy and momentum incorporate the oscillating motion of the particles, and can sometimes form a closed system with the nonresonant particles. 
	}
	\label{fig:subsystems}
\end{figure}

These two conservation relations, summarized in Fig.~\ref{fig:subsystems}, will allow for a very clear interpretation of the evolution of the bump-on-tail instability, which will guide our subsequent analysis of the alpha channeling problem.
For a more detailed discussion, see Ref.\cite{Ochs2021WaveDriven}.

\section{Bump-On-Tail Instability and Self-Consistent Quasilinear Forces} \label{sec:btiQL}

The detailed kinetic mathematics of the bump-on-tail instability are worked out in several good textbooks \cite{kralltrivelpiece,Tsytovich1977,davidson1973methods}; here, we will simply give a brief overview of the relevant features of the problem.

\begin{figure}[t]
	\center
	\includegraphics[width=0.9\linewidth]{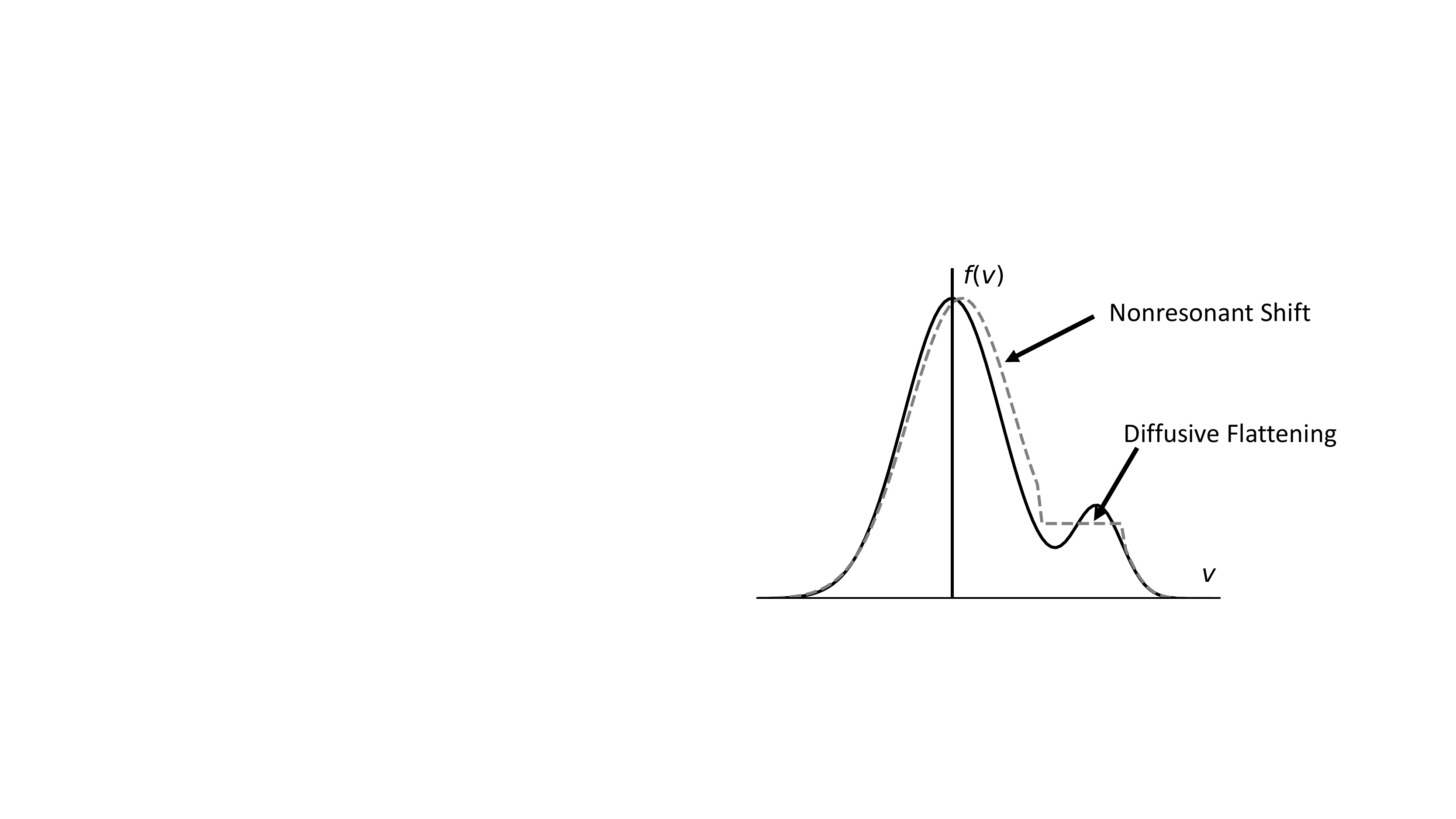}
	\caption{
		Quasilinear evolution of the bump-on-tail instability.
		Resonant particles lose momentum, which shows up as Minkowski momentum in the wave.
		However, the wave has no electromagnetic momentum, and so the 
	}
	\label{fig:bti}
\end{figure}

The bump-on-tail instability is an electrostatic kinetic instability in a homogeneous, unmagnetized plasma.
It occurs for electron Langmuir waves, with the ions taken as a neutralizing background.
The instability is triggered when, near the phase velocity $v_p = \omega_r/k_r$, the particle distribution satisfies $v \paf{f}{v} > 0$; i.e.~when there are more particles at high energy than low energy (black line, Fig.~\ref{fig:bti}).
Then, the wave-induced diffusion flattens out the bump, transferring energy and momentum out of the resonant particles.
Keeping in mind our conservation relation from Eq.~(\ref{eq:cons1DMRP}), this energy and momentum appear as Minkowski momentum in the wave.

If one just looked at the resonant electrons, it would thus look as if the plasma was losing energy and momentum.
However, from Eq.~(\ref{eq:cons1DEMP}), we know that the total energy and momentum in the plasma must remain constant, since the wave has no electostatic momentum.
Thus, the nonresonant particle distribution must shift, canceling the momentum loss from the resonant particles, to satisfy both momentum conservation equations---and this is exactly what happens (Fig.~\ref{fig:bti}).

It turns out that we can actually derive a very simple expression\cite{Ochs2020,ochs2021nonresonant} for the quasilinear force on the plasma that demonstrates this result, and generalizes to any electrostatic wave.
We start with the linear theory, which for an electrostatic wave begins with the Poisson equation:
\begin{align}
	-\nabla^2 \phi &= 4 \pi \sum_s q_s n_s, \label{eq:poisson}
\end{align}
Fourier transforming gives the general form of dispersion relation for an electrostatic wave:
\begin{align}
	0 &= D \equiv 1 + \sum_s D_s; \quad D_s \equiv -\frac{4 \pi q_s }{k^2} \frac{\tn_s}{\tphi}. \label{eq:ESDispersionGeneral}
\end{align}
Usually, we get the real and imaginary frequencies by Taylor expanding this dispersion relation in small $|\omega_i| / |\omega_r|$, yielding real and imaginary components of the dispersion relation:
\begin{align}
	0&=1 + \sum_s D_{rs} \label{eq:realDisp} \\
	0&= \sum_s  \lp i \omega_i \pa{D_{rs}}{\omega_r} + i D_{is} \rp \label{eq:imagDisp},
\end{align}
Here, $D_{is}$ and $D_{rs}$ are the real and imaginary parts of $D_s$ when evaluated at real $\omega,k$.

Now, we can get the average force on each species over a wave cycle by taking the average of the field-density correlation:
\begin{align}
	F_s^i &= q_s \llangle n_s E^i \rrangle = \frac{q_s}{2} \Re \left[\tE^{i*} \tn_s \right] \\
	&= 2 W_{EM} k^i \left[ D_{is} +  \omega_i \pa{ D_{rs}}{\omega_r} \right]. \label{eq:QLforce1D}
\end{align}
Here, the second line used the definition of $D_s$ and $\tE^i = -ik^i\tphi$.

Note that Eq.~(\ref{eq:QLforce1D}) automatically satisfies momentum conservation, since if we sum over all species, the term in brackets is simply the imaginary part of the dispersion function (Eq.~(\ref{eq:imagDisp})), which must by definition be 0.
As explained in more detail in Refs.\cite{Ochs2020,ochs2021nonresonant,Ochs2021WaveDriven}, the first term represents the resonant diffusion (and can be shown to be consistent with the absorption of Minkowski momentum\cite{Ochs2021WaveDriven}), and the second term is the nonresonant recoil response. 
Thus, the nonresonant recoil naturally arises in this momentum-conserving quasilinear framework, which forms the basis for our subsequent analysis.

\section{Current Drive by Wave-Mediated Momentum Exchange} \label{sec:currentDrive}

The quasilinear theory of the Section~\ref{sec:btiQL} leads to an important and immediate result in current drive, which we explore in this section.
This section can thus be skipped by readers interested only in questions of cross-field charge transport and rotation drive.

Consider again the Langmuir wave from Section~\ref{sec:btiQL}, but now without the bump-on-tail, so that the wave damps rather than amplifies.
Because of the lack of momentum in a purely electrostatic planar field, equal and opposite momenta, and thus equal and opposite currents, will be driven in the resonant and nonresonant electron populations.
To get net current, therefore, some of this momentum has to be transferred into the ions.
Such momentum exchange can be provided by collisions.
For instance, since Langmuir waves have a high phase velocity relative to the electron thermal velocity, and because the Coulomb collision frequency with the background ions scales as $v^{-3}$, the resonant current driven in the tail electrons will be much longer-lived than the nonresonant current.
Thus, a net current is produced on collisional timescales \cite{fisch1978currentDrive,fisch1980creating,stix1992waves,bellan2008fundamentals}.

However, such a situation spells trouble for current drive by other waves, which don't exhibit collisional timescale separation between resonant and nonresonant particles.
For instance, ion-acoustic waves (IAWs) have a phase velocity $v_{p} \sim C_s \approx \sqrt{Z T_e/m_i} \ll v_{the}$, so that resonant electrons are right in the middle of the thermal distribution, and thus lack a collisional timescale separation from the nonresonant particles.
One might then think that it is impossible to drive currents with IAWs.

However, there is a major difference between the IAW and the electron Langmuir wave: both ions and electrons participate in the IAW.
Thus, momentum conservation alone does not say where the nonresonant recoil will go; it could go either to electrons or ions.
If the recoil goes to the ions, it will drive negligible current, leaving uncompensated resonant electron current, and thus net current drive.

Our quasilinear force from Eq.~(\ref{eq:QLforce1D}) can quickly tell us where this recoil goes.
Combining this equation with the definition of the Minkowski momentum from Eqs.~(\ref{eq:wpMinkowski}-\ref{eq:actionES}), we find the very compact expression\cite{Ochs2020}:
\begin{align}
	\frac{dj^i}{dt} &= \sum_s \frac{q_s}{m_s} \frac{dp^i_s}{dt} = -\frac{dp^i_M}{dt} \sum_s \frac{q_s}{m_s} \lp \gammab_s  - \etab_{s} \rp  \label{eq:djdtdPdt}
\end{align}
where we have now defined the species resonant ($\gammab_s$) and nonresonant ($\etab_s$) response coefficients:
\begin{align}
	\gammab_s \equiv \frac{D_{is}}{\sum_{s'} D_{is'} }; \quad \etab_{s} \equiv \frac{\paf{D_{rs}}{\omega_r} }{\sum_{s'} \paf{D_{rs'}}{\omega_r} }. \label{eq:etab}
\end{align}
For a wave that damps on multiple species, these coefficients take a value between 0 and 1, representing the fraction of the resonant and nonresonant response that goes to each species.
For a typical, long-wavelength IAW, we have:
\begin{align}
	D_{ri} &= -\frac{\omega_{pi}^2}{\omega^2}; \quad D_{re} = \frac{1}{k^2 \lambda_{De}^2},
\end{align}
so that $\etab_i = 1$ and $\etab_e = 0$.
Thus, the entire nonresonant response is in the ions, and the driven resonant electron current is uncompensated.
This quick result is strengthened in Ref.\cite{Ochs2020}, where we show that for any IAW, the electron resonant current $\gammab_e$ is \emph{always} significantly larger than the electron nonresonant response $\etab_e$, even when taking into account next-order corrections.

Because this current drive mechanism relies on the momentumless wave field to rearrange momentum between the different plasma species, it has been termed "wave-mediated momentum exchange".
This method of current drive was first discovered using a much more complex kinetic calculation by Manheimer\cite{manheimer1977iawCurrent} for IAWs, and Kato\cite{kato1980electrostatic} for IAWs and LH waves; here, we see that it follows very straightforwardly from the form of the force in Eq.~(\ref{eq:QLforce1D}).

The case of the ion-acoustic wave is intriguing, because shocks, which are ubiquitous in astrophysical systems, tend to produces ion-acoustic waves in their wake \cite{sagdeev1963shock,chen2012introduction,taylor1970IAWshock}.
Thus, the same shocked systems that are predicted to produce magnetic fields in astrophysics via the Biermann battery \cite{zweibel2013seeds,kulsrud1997magnetogenesis,gnedin2000magnetogenesisBiermann,hanayama2005biermann,hanayama2006snr,naoz2013biermannMagnetogenesis,biermann1950battery,Ridgers2021}, such as those around super-nova remnants \cite{hanayama2005biermann, hanayama2006snr}, could produce magnetic fields via IAW-driven currents, providing an alternate possible mechanism for the origin of cosmological seed magnetic fields.

This intriguing possibility is explored in Ref.~\cite{Ochs2020a}.
To calculate the magnetic field growth due to the IAW current, one must account for the self-inductance of the plasma that fights the creation of a magnetic field.
This can be accomplished, as in magnetohydrodynamics, by neglecting the electron inertia in the electron momentum equation.
Thus, the electron momentum equation must become force-free, which is enforced by the generation of an electric field.
If this electric field has curl, a magnetic field is generated.
The change in the magnetic field can be calculated by combining Faraday's Law and the electron momentum equation to obtain\cite{Ochs2020a}:
\begin{align}
	\pa{\ve{B}}{t} = \nabla \times (\vv \times \ve{B}) - \frac{c}{e} \nabla \times \lp \frac{\ve{F}_e}{n_e}  \rp. \label{eq:biermannInduction}
\end{align}
Here, $\ve{F}_e$ is the force on the electrons from everything other than the large-scale electric and magnetic fields.
If this force is the pressure force, one obtain the Biermann battery; if it is the quasilinear force from the IAW, one obtains a corresponding ``IAW Battery.''

\begin{figure}[t]
	\center
	\includegraphics[width=\linewidth]{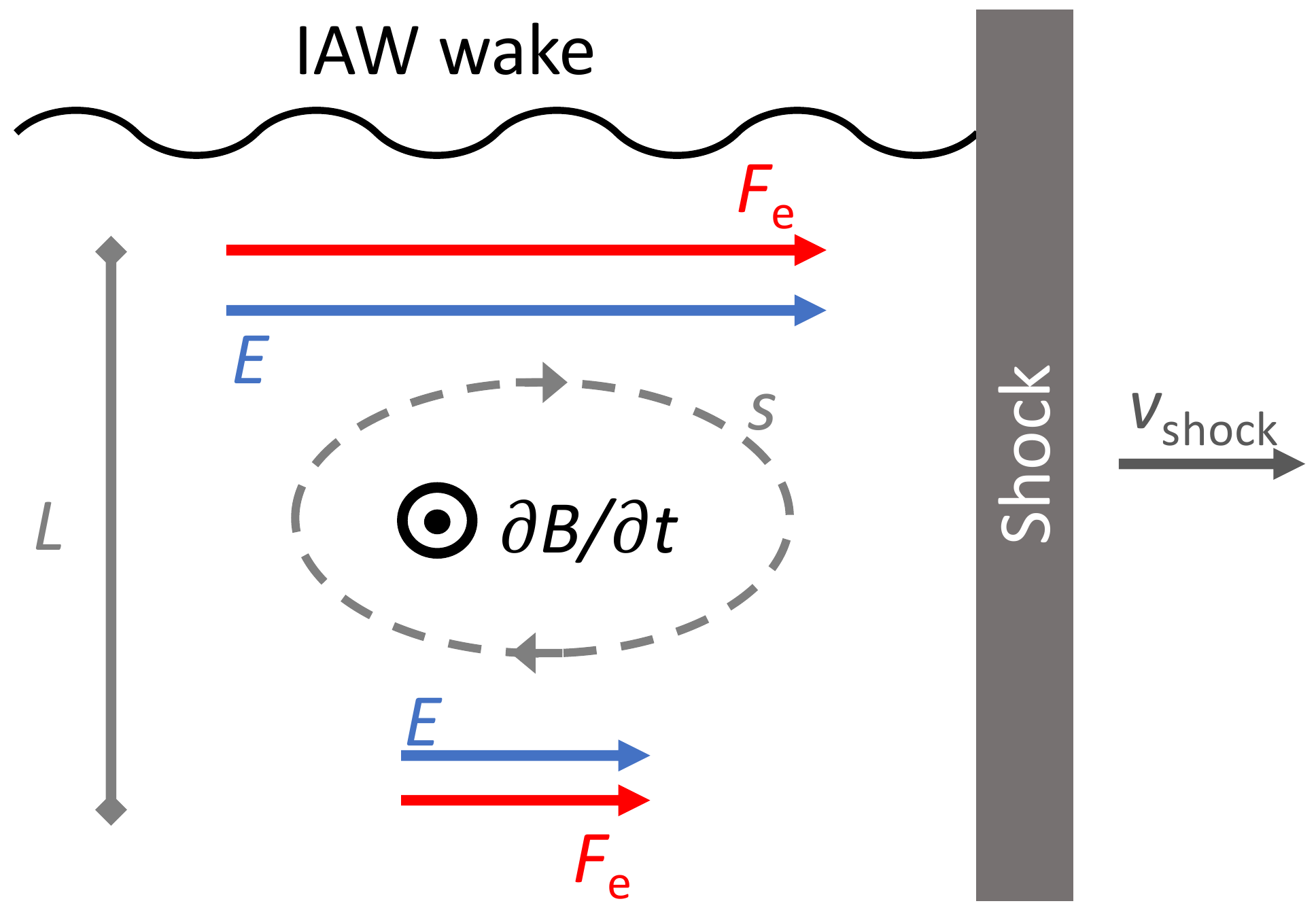}
	\caption{Mechanism of magnetogenesis by ion-acoustic waves.
		A shock propagates through the ISM, producing ion acoustic waves in its wake.
		These waves produce a force $\ve{F}_{e}$ electrons,  inhomogeneous on a scale $L$ in the ISM.
		A compensating inhomogeneous electric field $\ve{E}$ arises to cancel this force.
		This field has curl (consider the integral of the field over the loop $s$), and thus induces a magnetic field $\ve{B}$.
		Figure from Ref.~\cite{Ochs2020a}.
	}
	\label{fig:magnetogenesisMechanism}
\end{figure}

A possible scenario for magnetogenesis by this IAW battery is shown in Fig.~\ref{fig:magnetogenesisMechanism}.
A shock from a super nova expands into the surrounding intersteller medium (ISM), encountering inhomogeneities in the medium.
These inhomogeneities lead in turn to inhomogeneities in the wave strength, and thus in the force experienced by the electrons.
A curl in this force results in a compensating curl in the electric field to keep the electrons force-free, and thus to the generation of a magnetic field.
Depending on the wave parameters, this process can sometimes produce fields faster than the Biermann battery mechanism \cite{Ochs2020a}.

\section{Linear and Quasilinear Theory of Alpha Channeling} \label{sec:alphaIvp}

Now we return to the subject of rotation drive by alpha channeling.
The case of the bump-on-tail instability in Section~\ref{sec:btiQL} suggests that the nonresonant particle will gain any momentum lost by the resonant alpha particles.
The analysis of Section~\ref{sec:momentumMagnetized}, meanwhile, suggests that in the absence of net momentum input, no charge transport is possible.

To see how this plays out in the case of alpha channeling by a plane wave, we would like to make use of our simple quasilinear theory from Eq.~(\ref{eq:QLforce1D}).
However, making use of this theory requires a linear theory, which until recently did not exist for alpha channeling.

\subsection{Historical Results and Challenges}

The primary reason that this linear theory has proved elusive is that the pole associated with the perpendicular Landau resonance disappears from the dispersion relation as soon as there is any magnetic field in the plasma.
This puzzling disappearance of Landau damping as soon as a plasma is even infinitesimally magnetized has been termed the ``Bernstein-Landau paradox,'' since it involves a discrepancy between the Bernstein (magnetized) and Landau (unmagnetized) kinetic dispersion relations \cite{Sukhorukov1997,charles2021bernsteinLandau}.

The Bernstein-Landau paradox is a mathematical artifact more than anything else.
It arises from the fact that, in calculating the forces in the linear magnetized theory, the force is always evaluated at the unperturbed particle position.
Because of this, a particle never forgets its phase with respect to the wave as it goes around repeatedly on its gyro-orbit, and so unless there is a resonance between the gyroperiod and the wave period, the particle will gain and lose equal amounts of energy in a periodic oscillation.
In the language of dynamical systems, the particle is trapped on a small cycle, or ``island,'' in phase space \cite{karney1979stochastic}, and thus only oscillatory motion is possible.

Diffusion, i.e. stochastic motion of the particle through phase space, can only occur if these islands are destroyed.
While it is true that, in sufficiently magnetized plasmas, some particles will remain trapped in these islands, there are a multitude of mechanisms which can allow loss of phase memory and stochastic diffusion in a physical plasma.
This can occur when there are multiple overlapping waves whose islands overlap (the ``Chirikov criterion'')\cite{Chirikov1969StochasticityReport,Chirikov1971Stochasticity}, when random collisions introduce sufficient stochasticity\cite{swanson1989plasma}, or when a single wave has a large enough amplitude to trigger nonlinear physics\cite{Karney1978,karney1979stochastic}.

This last case was examined in full nonlinear detail by Karney, who derived a diffusion coefficient for the resonant particles in perpendicular velocity space.
To derive this diffusion coefficient, Karney solved for the dynamics of a particle undergoing Larmor rotation in the presence of an electrostatic wave of arbitrary strength.
Because the nonlinear mathematics involved were very complicated, the theory was limited to planar waves with constant amplitude in time---precluding a self-consistent quasilinear theory.
However, the theory did give a diffusion coefficient for the resonant particles, which can be written in perpendicular energy space as\cite{Ochs2020}:
\begin{align}
	D^{KK} &\equiv \frac{m_\alpha^2}{2} \lp \frac{q_\alpha k \phi_a }{m_\alpha} \rp^2  \frac{v_p^2}{\sqrt{k^2 v_\perp^2 - \omega_r^2}} H\lp v_\perp -  v_p \rp,
	\label{eq:finalQLResonantGyroDiffusion}
\end{align}
where $\phi_a$ is the wave amplitude, $v_\perp = \sqrt{2K/m_\alpha}$, and $H(x)$ is the Heaviside function.

Using the existence of this diffusion, Fisch and Rax\cite{fisch1992current,Fisch1992} used the relation between momentum and gyrocenter to show that every time a particle receives a kick $\Delta K$ in perpendicular energy, it receives a correlated kick $\Delta \ve{X} / \Delta K = \ve{k} \times \hat{b} / m_\alpha \omega \Omega_\alpha$ in gyrocenter position.
Wave amplification occurs when there are more particles at high than low energy along this path, i.e. when:
\begin{align}
	\lp \pa{}{K} + \frac{\ve{k}\times \hat{b}}{m_\alpha \omega \Omega_\alpha} \cdot \pa{}{\ve{X}}\rp F_{\alpha 0} (K,\ve{X}) > 0, \label{eq:amplificationCondition2}
\end{align}
where $F_{\alpha 0} (K,\ve{X})$ is the lowest-order distribution function for the resonant alpha particles.

As we move to develop a self-consistent linear and quasilinear theory of alpha channeling, our main targets are the diffusion coefficient in Eq.~(\ref{eq:finalQLResonantGyroDiffusion}) and the amplification condition in Eq.~(\ref{eq:amplificationCondition2}).

\subsection{Linear Theory}

The requisite self-consistent linear theory of alpha channeling, for a poloidal plane wave ($\ve{k} \parallel \hat{y}$), is developed in Ref.~\cite{ochs2021nonresonant}.
The basic idea is to construct a theory for the alpha particles in a magnetized plasma that contains the Landau resonances perpendicular to the magnetic field (Fig.~\ref{fig:alphaChannelingSchemIntro}) that underlie the diffusion process in alpha channeling.
Since these resonances exist in the \emph{unmagnetized} kinetic theory, our approach is to transform the unmagnetized kinetic theory, which lives naturally in local phase space coordinates $x^i \equiv f_\alpha(x,y,v_x,v_y)$, to gyrocenter-energy coordinates $X^i \equiv F_\alpha(X,Y,K,\theta)$.
To model the nonlinear loss of gyrophase structure, we force the zeroth-order alpha particle distribution function to be independent of gyrophase $\theta$, and average all equations over $\theta$.
Indeed, a similar approach has been used before in homogeneous plasmas to calculate damping on alpha particles\cite{swanson1989plasma,Barbato1991,Barbato2004}; in contrast, here we allow the distribution function to be inhomogeneous along $X$, i.e. to have a radial alpha particle gradient.

\begin{figure}[t]
	\center
	\includegraphics[width=\linewidth]{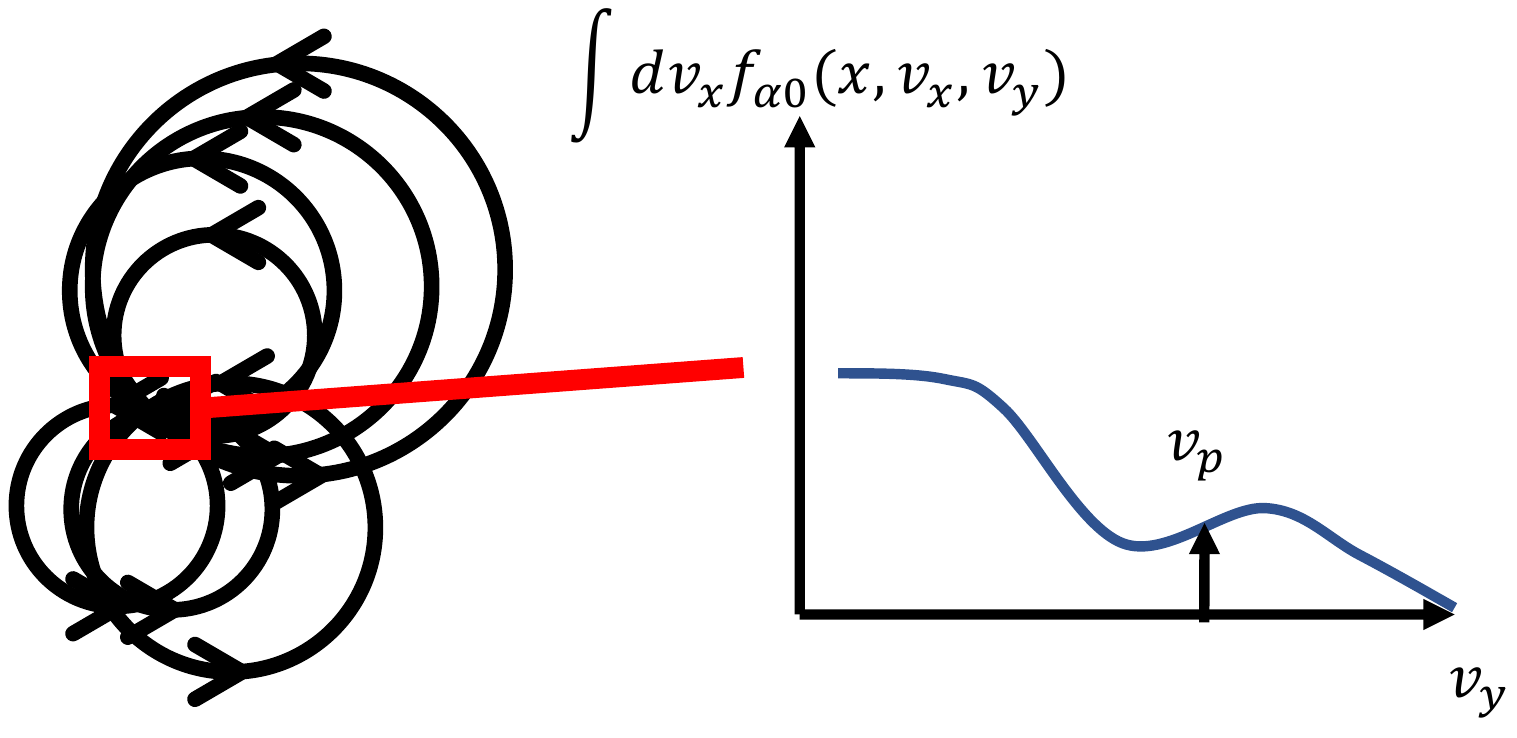}
	\caption{Equivalence of alpha channeling and the bump-on-tail instability.
		If a distribution $F_{\alpha 0} (X,K)$ for which alpha channeling occurs, i.e. for which $\omega_i>0$ in Eq.~(\ref{eq:channelingOmegaI}), is examined in local coordinates, then the local distribution $f_{\alpha 0}$ exhibits a bump-on-tail instability, where $\sgn(v_p) \int dv_x \paf{f}{v_y} > 0$.
	}
	\label{fig:btiAc}
\end{figure}

Treating the ions and electrons as cold fluids in the LH frequency range $\Omega_i \ll \omega_{LH} \ll \Omega_e$, we find that the alpha particles introduce an imaginary component to the dispersion relation, resulting in an imaginary frequency\cite{ochs2021nonresonant}:
\begin{align}
	\omega_i &= \frac{\pi}{2}   \frac{\omega_{p\alpha}^2}{\omega_{pi}^2} \frac{\omega_{LH}^4}{|k_y|^3} m_\alpha^2 \int dv_x   \lp \pa{F_{\alpha0}}{K}  + \frac{k_y}{m_\alpha \omega_r \Omega_\alpha} \pa{F_{\alpha0}}{X}\rp_{X^*,K^*}, \label{eq:channelingOmegaI}
\end{align}
where $X^* \equiv x+v_p/ \Omega_a, K^*\equiv  m \lp v_x^2 + v_p^2  \rp/2$.
The term in parenthesis can be recognized as precisely the alpha channeling amplification condition from Eq.~(\ref{eq:amplificationCondition2}), but now weighted properly over a distribution of alpha particles, rather than simply a single particle.
Eq.~(\ref{eq:channelingOmegaI}) represents the first calculation of a linear amplification rate from the alpha channeling process.

The amplification condition for LH wave alpha channeling was originally derived by demanding that the gyrocenter distribution of alpha particles have more particles at high energy than low energy along the gyrocenter diffusion path. 
In contrast, Eq.~(\ref{eq:channelingOmegaI}) is just a coordinate transform of the imaginary frequency result from Landau damping or the bump-on-instability.
Nevertheless, these two very different approaches lead to the same amplification condition, and thus, evidently, represent the same physical process.
In other words, if we take a distribution function $F_{\alpha 0} (X,K)$ in gyrocenter coordinates for which alpha channeling occurs, and we look at it locally in physical coordinates $f_{\alpha 0}(x,v_x,v_y)$, we will find a bump-on-tail instability along $v_y$ (Fig.~\ref{fig:btiAc}).
Establishing the equivalence of these two instabilities, which were thought to be entirely distinct, is a rewarding result of developing the linear theory.

\subsection{Quasilinear Theory}

Just as transforming the unmagnetized linear kinetic theory to gyrocenter coordinates yielded the alpha channeling amplification condition, transforming the unmagnetized quasilinear kinetic theory yields the diffusion equation for the resonant alpha particle distribution\cite{ochs2021nonresonant}:
\begin{align}
	\llangle \pa{F_{\alpha 0}}{t} \rrangle_t  &= \frac{d}{dK}\biggr|_\text{path} \left[D^{KK} \frac{d}{dK}\biggr|_\text{path} F_{\alpha 0} \right] \label{eq:ChannelingPathDiffusionEquation}\\
	\frac{d}{dK}\biggr|_\text{path} &\equiv \lp \pa{}{K} + \frac{k_y}{m_s \omega_r \Omega_\alpha} \pa{}{X}\rp.
\end{align}
Thus, the theory recovers diffusion along the channeling diffusion path, definitively establishing that the theory captures the alpha channeling process.

\begin{figure}[t]
	\center
	\includegraphics[width=\linewidth]{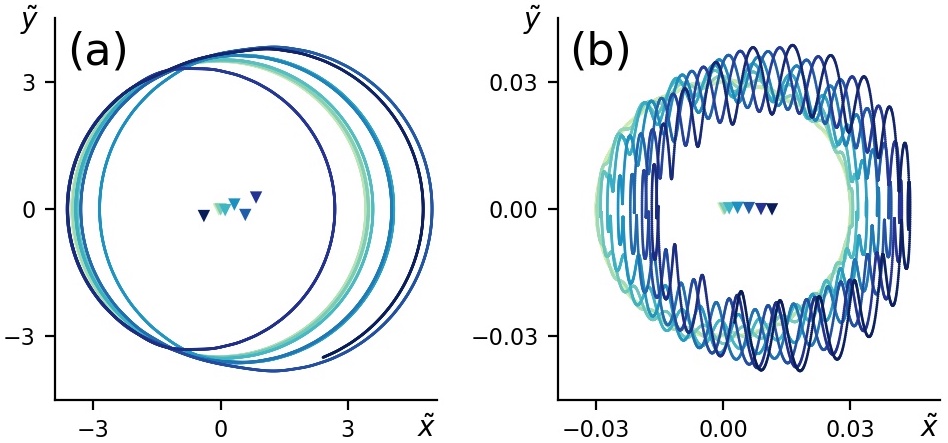}
	\caption{Simulated single-particle trajectories in the $x$-$y$ plane of (a) hot particles ($v_0 = 3.5 v_p$) and (b) cold particles ($v_0 = 0.03 v_p$) in a growing electrostatic wave.
		Lines are particle positions, and triangles are orbit-averaged gyrocenter positions.
		Axes are normalized to $\rho_{ps} \equiv \omega / k \Omega_s$, i.e. $\tilde{x}=x/\rho_{ps}$ and $\tilde{y}=y/\rho_{ps}$.
		The color indicates time, light to dark.
		The hot particles diffuse stochastically.
		The cold particles exhibit a clear shift in gyrocenter toward positive $\tilde{x}$, which ultimately cancels the charge transport from the hot alpha particles.
		Figure from Ref.~\cite{ochs2021nonresonant}.
	}
	\label{fig:particleOrbits}
\end{figure}

\begin{figure}[t]
	\center
	\includegraphics[width=\linewidth]{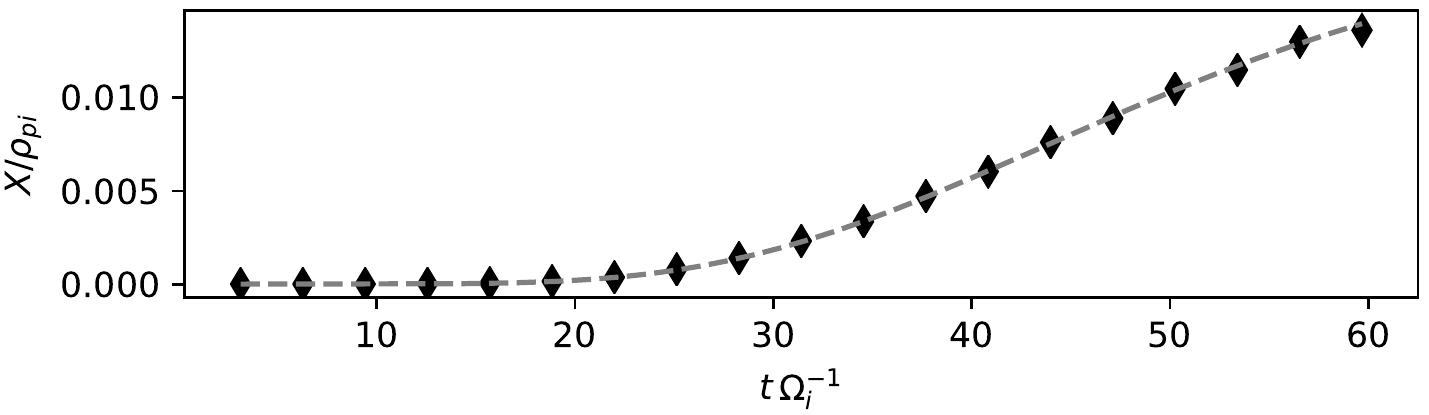}
	\caption{Change in gyrocenter position $X$ for the particle in Fig.~\ref{fig:particleOrbits}b due to the slow ramp-up of the electrostatic wave.
		The gyro-period-averaged position in the simulations (filled black diamonds) agrees well with the predicted shift\cite{ochs2021nonresonant} (gray dashed line) due to the nonresonant recoil.
		Figure from Ref.~\cite{ochs2021nonresonant}.
	}
	\label{fig:gyrocenterShift}
\end{figure}

Since the new linear theory captures the alpha particle response as an imaginary component of the dispersion relation $D_{i\alpha}$, we can apply the simple quasilinear theory force theory from Eq.~(\ref{eq:QLforce1D}).
Since for the bulk components we have $D_{re} = \omega_{pe}^2/\Omega_e^2$, $D_{ri} = -\omega_{pi}^2/\omega^2$, the recoil force (as in the case of ion acoustic waves in Section~\ref{sec:currentDrive}) will be entirely in the ions, and lead to an $\ve{F} \times \ve{B}$ drift (and corresponding charge transport) that entirely cancels the current from the resonant alpha particles.
The presence of this drift in a time-growing plane wave is confirmed\cite{ochs2021nonresonant} by single-particle simulations (Fig.~\ref{fig:particleOrbits}) of the full Lorentz force in a growing high-frequency electrostatic wave, which agree well with the predictions of quasilinear theory (Fig.~\ref{fig:gyrocenterShift}).

As a result of this analysis, we see is that charge cannot be extracted, and thus $\ve{E} \times \ve{B}$ rotation cannot be driven, by a purely poloidal plane wave that grows in time.

\section{Spatially Structured Waves} \label{sec:alphaBvp}

However, this conclusion is not the whole story.
Often, in problems of plasma control, we drive a wave from the outer radial plasma boundary, and are interested in what this wave looks like in steady state.
Understanding what charge transport looks like for this state requires analyzing waves with spatial structure.
For simplicity, we will consider absorption of a damped wave; however, the core results apply also to amplification of a wave due to alpha channeling.

\begin{figure}[tp]
	\center
	\includegraphics[width=\linewidth]{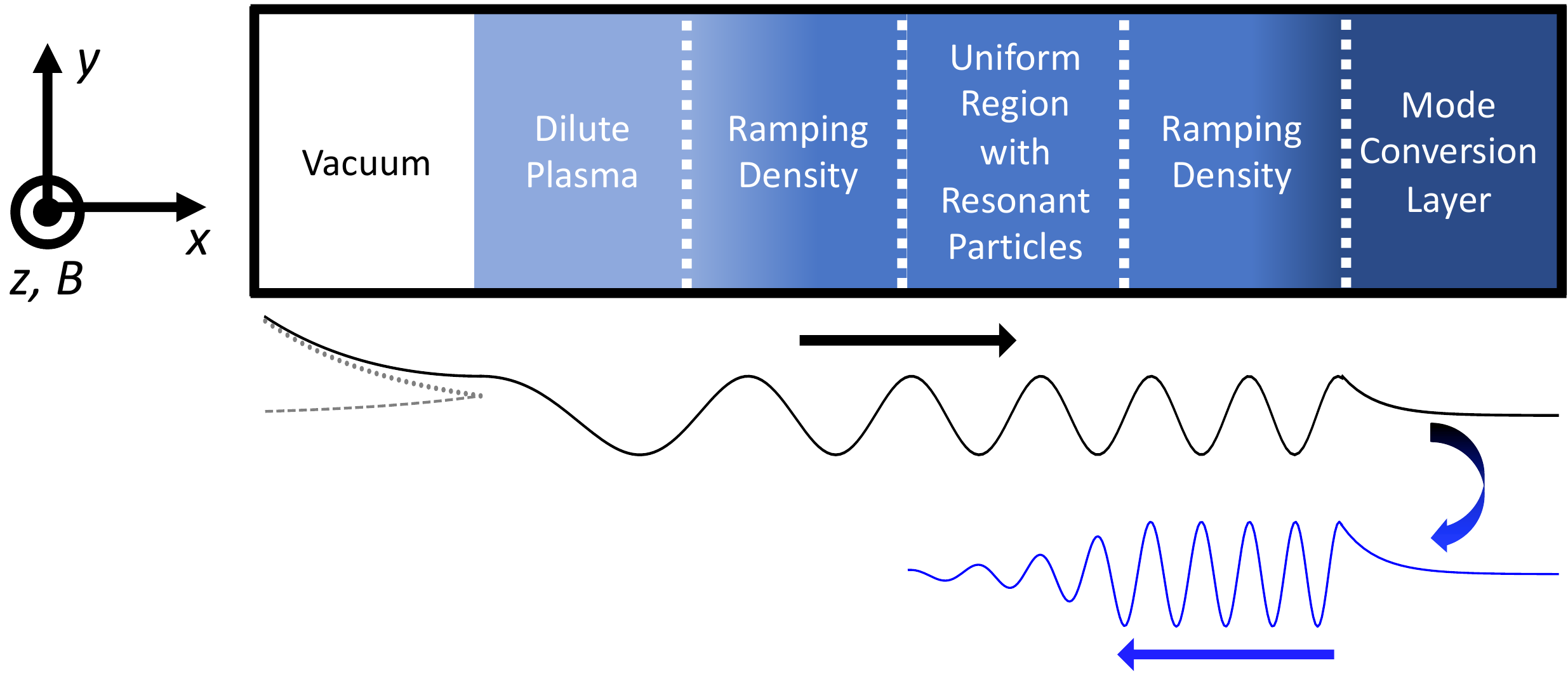}
	\caption{
		Model for the structured wave analysis from Ref.~\cite{Ochs2021WaveDriven}.
		Darker color represents higher density.
		A vacuum evanescent wave (black), consisting of a forward-decaying ``incident'' wave (dotted gray) and backward-decaying ``reflected'' wave (dashed gray), converts to a ``transmitted'' slow wave at the plasma / vacuum interface.
		This slow wave then continues to propagate inward as the plasma density increases until it hits the lower hybrid resonance, where it mode converts into an electrostatic lower hybrid wave (blue).
		Wave action is conserved during the in-plasma propagation and mode conversion.
		This wave then propagates back to a uniform region containing resonant particles, where it damps.
		Figure from Ref.~\cite{Ochs2021WaveDriven}.
	}
	\label{fig:waveInjectionModel}
\end{figure}

To look at this new situation, we will consider the model from Ref.~\cite{Ochs2021WaveDriven}, shown in Fig.~\ref{fig:waveInjectionModel}.
This simple model is motivated by the coupling of waveguides for lower hybrid current drive \cite{Brambilla1976,Brambilla1979}.
For such waves, an evanescent wave in a vacuum region converts into a plasma slow wave as the plasma density ramps up past the point where $|\omega_{pe}| > |\omega|$.
It then propagates up to the mode-conversion layer, where it reflects as an electrostatic lower hybrid wave \cite{Stix1965}.
We make two main simplifications to dramatically reduce the problem complexity.
First, we assume all damping occurs in a uniform region with resonant particles, so that we do not have to account for changing wave parameters in the resonant region.
Second we take the plasma to have a sharp boundary, where the evanescent wave converts to a slow wave.
This reduces the coupling calculation to a Fresnel-style\cite{Griffiths2017,Jackson1999} boundary-matching condition, dramatically simplifying the mathematics.

\subsection{Conservation Relations Revisited}

The conservation relations from Section~\ref{sec:conservation1D} relax considerably in a plasma with spatial structure.
The general conservation relations are best expressed in terms of the energy-momentum tensor (EMT) of the system:
\begin{align}
	\ve{T} &= \bvec W & \ve{S}/c \\ c \ve{p} & \boldsymbol{\Pi} \evec.
\end{align}
In a closed system, conservation laws follow from the vanishing of the 4-divergence of this tensor, expressed in Cartesian coordinates in flat spacetime as:
\begin{align}
	\frac{1}{c} \pa{}{t} T^{\mu 0} + \pa{}{x_i} T^{\mu i} &= 0. \label{eq:stressTensorDiv0}
\end{align}
Energy conservation corresponds to the $\mu = 0$ portion of this equation, and momentum conservation to the $\mu =  1-3$ components.

Looking at momentum conservation, we are primarily interested in the appearance of the stress terms $\boldsymbol{\Pi}$:
\begin{align}
	\Pi_{EM}^{ij} &= -\frac{1}{4\pi} \lp E^i E^j - \frac{1}{2} \delta^{ij} E^2 +  B^i B^j - \frac{1}{2} \delta^{ij} B^2 \rp \label{eq:piEM}\\
	\Pi_{P}^{ij} &= \int  m v^{i} v^j f (t,\ve{x},\ve{v}) d \ve{v} \label{eq:piP}\\
	\Pi^{ij}_M &= p^i_M v_g^j, \label{eq:piMinkowski}
\end{align}
where we have introduced the group velocity of the wave:
\begin{align}
	v_g^i &= -\frac{\paf{D_r}{k_{ri}}}{\paf{D_r}{\omega_r}}. \label{eq:groupVelocity}
\end{align}

For the purposes of charge transport, we are interested in the force along the symmetry direction $y$.
As discussed in Ref.~\cite{Ochs2021WaveDriven}, for the purposes of this calculation, the tensors $(\ve{T}_{EM} + \ve{T}_{P})$ and $(\ve{T}_{M} + \ve{T}_{RP})$ both represent closed systems---a fact we make extensive use of in the ensuing analysis.

\subsection{Local Momentum Conservation}

In the presence of a wave, the kinetic stress tensor gives rise to a stress component due to the oscillating motion of the particles, known as the Reynolds stress:
\begin{align}
	\Pi^{ij}_\text{Rey} &= m n_0 \llangle u_1^i u_1^j \rrangle = \frac{1}{2} m n_0 \Re \left[ \tu_a^i \tu_a^j\right] ,\label{eq:ReynoldsStress}
\end{align}
where $\ve{\tu}_1$ is the local complex velocity amplitude of the wave.
This Reynolds stress plays an important role in calculating turbulent forces on a plasma \cite{Diamond1991,Berry1999,Jaeger2000,Myra2000,Myra2002,Myra2004,Diamond2008}.
The presence of the Maxwell stress tensor and the Reynolds stress dramatically change the conservation calculations.
As shown in Ref.~\cite{Ochs2021WaveDriven}, for the lower hybrid wave, the two combine in just such a way as to cancel the nonresonant (recoil) force on the plasma, leaving only the resonant force, in a way consistent with momentum conservation.
Thus, in the spatially-structured, steady-state problem of alpha channeling, resonant particles \emph{can} be extracted, and $\ve{E} \times \ve{B}$ rotation \emph{can} be driven, thanks to the torque exerted on the plasma by the Maxwell and Reynolds stresses.

\subsection{Global Momentum Conservation}

One might still wonder where this torque is coming from; is there an externally-applied net torque on the plasma consistent with this force on the resonant alpha particles? 
If not, then this local force on the plasma must be compensated somewhere else within the plasma, leading to momentum rearrangement within the plasma (and thus the generation of shear flow), but no net spin-up of the plasma.

To look at this question, we turn in Ref.~\cite{Ochs2021WaveDriven} to a Fresnel model of the wave transition between the vacuum and the plasma.
Now, the evanescent wave in the vacuum has a well defined electromagnetic momentum flux (Maxwell stress), but no well-defined Minkowski flux.
However, it is possible to show that, for a lower hybrid wave, the $x$-directed flux of $y$-directed \emph{electromagnetic} momentum $\Pi_{EM}^{yx}$ through the vacuum, is equal to the $x$ directed flux of $y$-directed \emph{Minkowski} momentum $\Pi_{M}^{yx}$ within the plasma.
Because the Minkowski EMT forms a closed system with the resonant particle EMT, this momentum all ends up in the resonant particles.
Thus, the force on the resonant particles that allows charge extraction is ultimately provided by the electromagnetic fields propagating through the vacuum at the plasma edge.
As a result, the entire plasma spins up due to the interaction with the wave, in a way consistent with local and global momentum conservation.

In the case of alpha channeling, this analysis suggests that electromagnetic energy will flow from the plasma into the antenna as a result of alpha channeling.
This raises the intriguing possibility of performing direct conversion of alpha particle energy into electrical energy at the antenna, which could be a much more efficient way to usefully harvest the energy than any thermal process.

\begin{figure}[t]
	\center
	\includegraphics[width=0.8\linewidth]{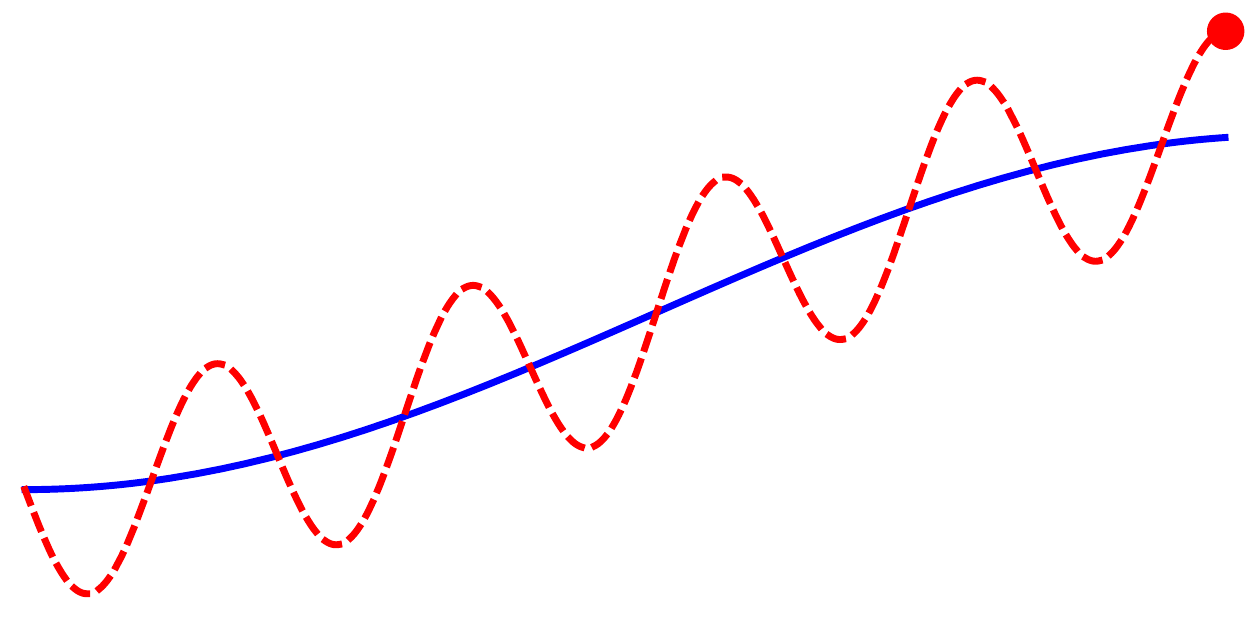}
	\caption{
		Oscillation center theories track the slow dynamics of the central position around which a particle quickly oscillates in a wave field.
	}
	\label{fig:ocSchem}
\end{figure}

\section{Oscillation Center Ponderomotive Theories} \label{sec:ocComp}

Although the above theory holds together well, and explains how charge extraction is consistent with momentum conservation, there are other, completely different approaches to calculating ponderomotive forces in the literature, which focus on the motion of the oscillation center.
It is important to see whether the theories are compatible and lead to the same conclusions.
Thus, in Chapter~7 of Ref.~\cite{Ochs2022Thesis}, we performed a detailed comparison of the two types of theory.

There are actually (at least) two distinct types of oscillation center theory in the literature: one \cite{Dodin2014Variational} which comes from separating out the quiver motion of a particle in the wave Lagrangian, and one, exemplified by Dewar\cite{Dewar1973}, which comes from applying a non-secular, near-identity canonical transform to the particle coordinates in the wave-particle interaction.
Each of these theories has somewhat different features, as we explore in Ref.~\cite{Ochs2022Thesis} for the simple case of an electrostatic wave in an unmagnetized plasma, which we summarize briefly here.

\subsection{Quiver Lagrangian}

We will start by examining the quiver-Lagrangian theory of the oscillation center\cite{Dodin2014Variational}.
Since it will already require a fair amount of complexity, we will consider only electrostatic waves in an unmagnetized plasma.
To get the Lagrangian that describes this theory, we separate out the oscillating motion of a particle from its smooth motion that changes on a longer timescale, obtaining:
\begin{align}
	\bar{\mathcal{L}} \approx \frac{1}{2} m  V^2  + \frac{1}{2} m \llangle \tv^2 \rrangle  + q \llangle \ve{\tx} \cdot \ve{E} (\ve{X},t) \rrangle. \label{eq:ocLagrangianMiddle}
\end{align}
Here, $\ve{X}$ and $\ve{V}$ represent the oscillation center position, and $\ve{\tx}$ and $\ve{\tv}$ describe the quiver motion, which can be solved simply at constant $\ve{X},\ve{V}$. 
Thus, the latter two terms combine to form a ponderomotive potential:
\begin{align}
	\Phi(\ve{X},\ve{V}) &= \frac{1}{4} \frac{q^2}{m} \frac{k^2 |\phi_a(\ve{X})|^2}{(\ve{k} \cdot \ve{V} - \omega)^2}, 
\end{align}
which gives the ponderomotive force via the Euler-Lagrange equations.

We show in Appendix~\ref{app:ponderomotive} that a naive application of this oscillation-center theory can yield erroneous conclusions about the ponderomotive force, but a more careful calculation agrees with the fluid theory.
As we show, forces in the two theories differ by three terms.
The first two---the Reynolds stress and the polarization stress---were identified by Gao\cite{Gao2007} in a cold fluid plasma.
We show that in a warm plasma, which is necessary wave packet propagation into an unmagnetized plasma or for purely-perpendicular modes in a magnetized plasma, there is also a third term of discrepancy, namely the kinetic stress of the oscillation centers themselves.
Only when accounting for these three terms do the theories agree, and support the conclusion that the nonresonant force along $y$ vanishes in steady state.

\begin{figure}[t]
	\center
	\includegraphics[width=\linewidth]{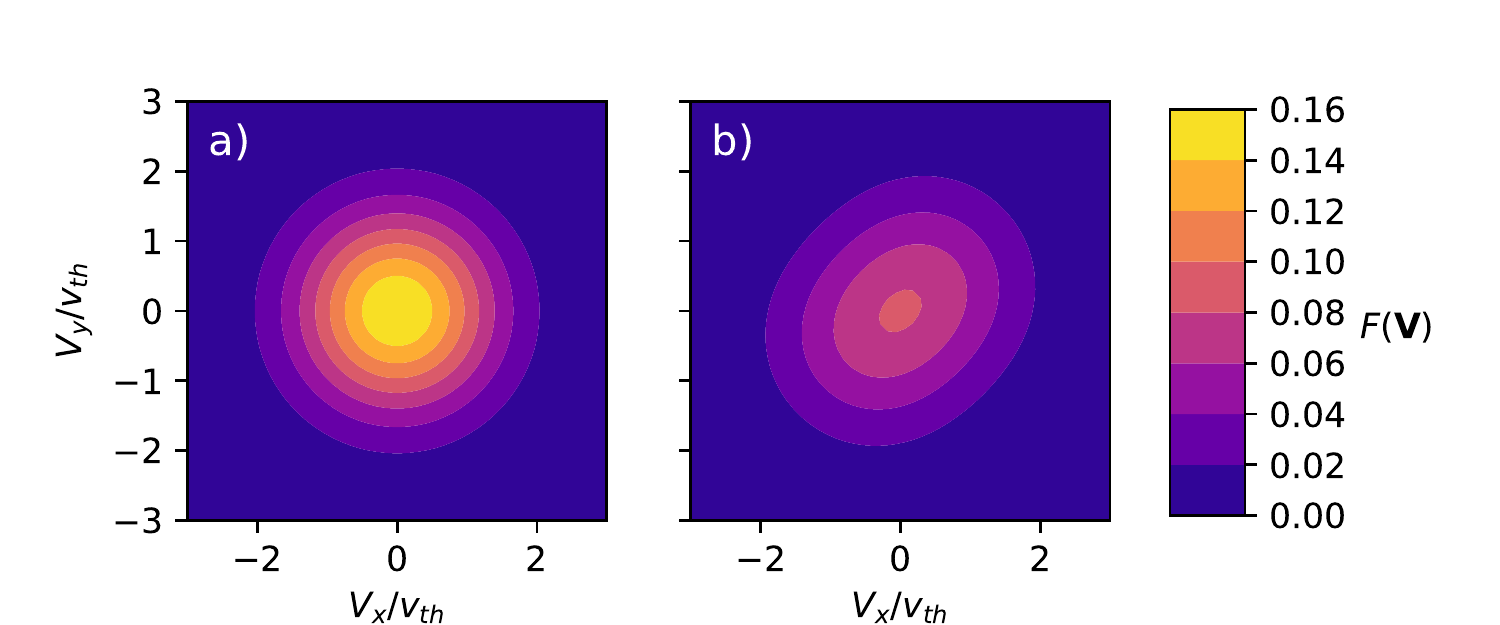}
	\caption{
		Oscillation center velocity distribution $F(\ve{V})$, as given by Eq.~(\ref{eq:ocFvFinal}), (a) far from the wave region, and (b) in the wave region.
		As explained in the appendix, the distribution in (b) is calculated for $W_{EM} / p_{th0} = 0.35$, $k_x v_{th} / \omega = k_y v_{th} / \omega = 1/4\sqrt{2}$.
		In the wave region, the distribution reduces in magnitude, and also stretches out along $\ve{k}$, resulting in anisotropy.
		This anisotropy provides the stress that allows a net force to be exerted on the plasma by the wave.
	}
	\label{fig:ocDistribution}
\end{figure}

\subsection{Near-Identity Transform}

The basic idea behind Dewar's theory is to use a canonical transformation, familiar from Hamilotonian mechanics \cite{Hand1998,Goldstein2001}, to transform from the physical particle coordinates $(\ve{x},\ve{p})$ to an oscillation center coordinate system $(\boldsymbol{\mathcal{X}},\boldsymbol{\mathcal{P}})$.
This coordinate system is chosen so as to only include the gradient ponderomotive potential and the resonant force, without any of the reactive time-dependent terms that can prove confusing.
To second order in wave amplitude, these coordinate systems are linked by the transformation \cite{Dewar1973}:
\begin{align}
	\dew{X} &= \ve{x} + \pa{S(\ve{x},\ve{p},t)}{\ve{p}} - \pa{S(\ve{x},\ve{p},t)}{\ve{x}} \cdot \frac{\partial^2 S(\ve{x},\ve{p},t)}{\partial \ve{p} \partial \ve{p} } \label{eq:Xdewar}\\
	\dew{P} &= \ve{p} - \pa{S(\ve{x},\ve{p},t)}{\ve{x}} + \pa{S(\ve{x},\ve{p},t)}{\ve{x}} \cdot \frac{\partial^2 S(\ve{x},\ve{p},t)}{\partial \ve{p} \partial \ve{x} } \label{eq:Pdewar}.
\end{align}
Here, $S$ is a quantity that defines the near-identity transformation, so we demand that $\llangle S \rrangle = 0$ at first order. 

In these coordinates, the ``force'' $d\dew{P}/dt$ from the waves on the particles consists of only two terms: the gradient of the ponderomotive potential, and the resonant diffusion \cite{Dewar1973}.
However, we have to be careful how we interpret this, because the ponderomotive force as we intuitively think of it cannot be simply identified with $d\dew{P}/dt$.
This is because, in general, what we think of intuitively as the ponderomotive force is either the change in the kinetic momentum $m \ve{V}$ of the oscillation center (in the quiver Lagrangian sense), or the change in the mean kinetic momentum $\llangle m n \ve{u} \rrangle$ of the plasma particles.
In contrast, $\dew{P}$ represents a canonical momentum with complex dependence on the ponderomotive potential, and with $\llangle \dew{P} \rrangle \neq \llangle \ve{p} \rrangle$, as evidenced by Eq.~(\ref{eq:Pdewar}).
Thus, while the Hamiltonian theories hold together theoretically, and elegantly conserve energy and momentum \cite{Dewar1973}, backing out the physical forces and currents from the solution in Hamiltonian phase space is a very complex task.

\subsection{Steady State} \label{sec:dewarSteadyState}

Nevertheless, Dewar's formulation is very useful for us, because it suggests something fundamental about ponderomotive forces from steady-state waves, and their effects on particle orbits.
In the absence of a wave, in a plasma confinement device, the orbit is determined by several constants of motion, including the particle energy and its momentum along the symmetry directions.
For instance, in a periodic cylinder, the orbit would be determined by the energy, and the momenta along the $\theta$ and $z$ directions.
Thus, the change in the orbit is determined by $\llangle dH/dt \rrangle$, $\llangle dp_\theta/dt \rrangle$, and $\llangle dp_z/dt \rrangle$.
However, in Dewar's theory, $\llangle H \rrangle = \llangle \mathcal{H} - K_2(S) \rrangle$, where $K_2$ is a ``renormalization energy'' that depends only on the function $S$.
For a steady-state wave, $\mathcal{H}$ is constant in time.
Furthermore, because $S$ is defined to not grow secularly, i.e. to be a fixed function of the wave amplitude, it is also constant in time.
Thus, $\llangle H \rrangle$ will be contant in time.
Similarly, looking at Eq.~(\ref{eq:Pdewar}), we see that $\llangle d\ve{p} /dt \rrangle$ is equal to $\llangle d\dew{P}/dt \rrangle$ plus terms that depend on $\llangle d S  / dt \rrangle$, so that by the same argument, in steady state, $\llangle \ve{p} \rrangle$ is a constant function of $\llangle \dew{P} \rrangle$.
Furthermore, $\llangle d\dew{P}/dt \rrangle = \llangle -d\mathcal{H}/d\dew{X} \rrangle = 0$ on a closed orbit.
Thus, Dewar's theory suggests that in steady state, nonresonant particles on closed orbits remain on those closed orbits, leaving only the resonant cross-field current.

In summary, Dewar's theory is not the best for understanding and calculating the reactive forces that are applied to nonresonant particles in time-growing waves.
However, once the consistency between the models has been established, it offers a deep explanation for the seemingly fortuitous disappearance of the nonresonant recoil force for steady-state, boundary-driven waves.

\section{Conclusions and Future Directions}

This paper reviewed recent work that addressed the question of whether alpha channeling, in transporting resonant ash across field lines, also transports net charge across field lines.  
Such charge transport would drive $\ve{E} \times \ve{B}$ rotation\cite{fetterman2008alpha,fetterman2010stationary,fetterman2012wave}, which can be advantageous for plasma stabilization, turbulence reduction, enhanced confinement in magnetic mirrors, and plasma centrifugation.  
In the end, we found that, at least for electrostatic waves, the answer depends on how the waves enter the plasma, with different results for plane waves that grow in time, as opposed to waves which enter from the plasma boundary.

To arrive at these results,
in Section~\ref{sec:momentumMagnetized}, we considered charge transport along $x$  due to collisions in a magnetized plasma slab, finding that it was very hard to move charge across field lines, since it required a net force to be applied along $y$.
In Section~\ref{sec:conservation1D}, we saw how this presented a problem for charge transport by a planar electrostatic wave, which contains no momentum.
In Section~\ref{sec:btiQL}, we saw how this momentum conservation played out in the bump-on-tail instability, motivating the development of a self-consistent quasilinear theory for alpha channeling.
After a foray into current drive theory in Section~\ref{sec:currentDrive}, in Section~\ref{sec:alphaIvp} we developed the first linear theory of alpha channeling as a prerequisite to evaluating the quasilinear forces on the plasma.
In so doing, we showed that alpha channeling is physically equivalent to the bump-on-tail instability, exposing a deep connection between two thought-to-be-distinct processes.
The resulting quasilinear theory allowed us to confirm the existence of the nonresonant recoil reaction in the case of a time-growing plane wave, which canceled the charge transport from the resonant ash, preventing the $\ve{E} \times \ve{B}$ rotation drive that was predicted by earlier theories.
As a silver lining, this return current was in the ions; thus, even though no rotation was driven, there was a different beneficial effect, in that for every alpha particle brought out by alpha channeling, two ions were brought into the core plasma, fueling the fusion reaction.

In Section~\ref{sec:alphaBvp}, we generalized the model to allow for spatially damped waves.
In this multi-dimensional problem, the Maxwell and Reynolds stresses played a prominent role, leading to the disappearance of the poloidal nonresonant force and its associated charge transport in steady state.
Thus, extraction of charge by a steady-state, boundary-driven wave was shown to be consistent with momentum conservation.
A Fresnel model of the wave propagation from vacuum to plasma, combined with a wave action conservation principle, confirmed that the momentum and energy required for this charge extraction were ultimately provided by electromagnetic energy and momentum flux in the evanescent wave traveling through the vacuum gap between the antenna and plasma.
Applied to alpha channeling, this analysis suggested the possibility of direct conversion of alpha particle energy into electric energy, as the alpha particle amplification results in an outwardly-directed Poynting flux, driving electromagnetic energy into the antenna.

Taken together, the analysis in Sections~\ref{sec:alphaIvp}-\ref{sec:alphaBvp} revealed a core difference between the plane-wave initial value problem of alpha channeling, where net charge cannot be extracted, and thus rotation cannot be driven, and the steady-state boundary-value problem, where it can be. 
In contrast to earlier models that focused only on the resonant particles, the analysis here respects the momentum conservation principles between the various plasma subsystems. 

In Section~\ref{sec:ocComp}, we compared our quasilinear model to the variational Lagrangian and Hamiltonian theories of the oscillation center, for the simple case of an electrostatic wave in an unmagnetized plasma.
We found that the resulting ponderomotive forces in the two models differed by the terms arising from the Reynolds, polarization, and oscillation center stresses, but ultimately agreed on the form of ponderomotive force.
Furthermore, an analysis of Dewar's near-identity transform theory suggested that the vanishing of the nonresonant recoil reaction for steady-state waves was likely to be a general result across many systems, at least for nonresonant particles on closed orbits in phase space.

This paper thus provides a cohesive and simple theory of rotation driven by alpha channeling, which demystifies how momentum conservation works out while allowing charge extraction.
Since it was intended to resolve a core mystery, our analysis throughout was deliberately simple, taking every simplifying assumption possible while retaining the core features of the problem.
Thus, there are many generalizations and extensions possible, which will be required in order to quantitatively connect to future experiments.

First, and likely most straightforward, is to extend the analysis from electrostatic waves to electromagnetic waves.
This is necessary to handle one of the systems originally envisioned by Fetterman, in which stationary magnetic perturbations at the plasma edge provide the moving wave in the rotating plasma frame \cite{fetterman2010stationary,fetterman2012wave}.
Although it requires us to move from a scalar to a vector theory, the essential features of the theory will be similar, but with the susceptibility in many cases taking over role of the dispersion function.
Of course, for an electromagnetic wave, the electromagnetic field does have momentum, somewhat altering the momentum balance and magnitude of the recoil in the plane wave initial value problem. 
However, the general principles from the oscillation center theories suggest that the core conclusions in the electromagnetic case will remain fairly similar: namely, there will be some canceling charge transport for a time-dependent plane wave due to the nonresonant recoil, but no recoil for the steady-state boundary-value problem, with momentum conservation in the latter case enforced by a combination of Reynolds and Maxwell stress. Thus, rotation drive will likely be possible for Fetterman's magnetostatic edge coil scheme\cite{fetterman2010stationary,fetterman2012wave} as well.

As we further generalize to consider the generation of shear flow in a plasma, the plasma must by definition become inhomogeneous, since the rotation profile is a function of radius.
This inhomogeneity will impact both the dissipation and the evolution of $\ve{k}$.
Handling this situation correctly will require integration of the momentum-conserving channeling theory with geometrical optics.

To fully solve for the rotation drive, it will be necessary to combine this improved theory with a theory of rotation dissipation.
This dissipation can be due to a variety of collisional processes\cite{stix1973decay,kolmes2019radial}, or ``anomolous'' transport such as that due to the stochasticity of the magnetic field lines\cite{finn1992particle,almagri1998momentum}.

Finally, in moving to more complex geometries, a variety of other considerations come into play.
In toroidal geometry, the waves can move particles between neoclassical orbits\cite{Herrmann1997} rather than gyrocenter orbits, opening the possibility for even larger rotation drive.
In mirrors, not only radial losses, but also losses from wave-induced scattering into the loss cone, must be taken into account.
Determining how these additional loss mechanisms affect the radial potential will require an involved, self-consistent, Pastukhov-style \cite{Pastukhov1974,McHarg1975,Cohen1978} analysis that accounts for the balance of ion and electron losses.
However, the dividends for such an analysis are potentially large, freeing open-field-line devices from the design constraints imposed by end electrodes.

Thus, although we have firmly established the theoretical basis for rotation drive by alpha channeling, we are really only at the beginning of exploring the details of how it can be used to tailor rotation profiles in various devices.

\section*{Acknowledgments}
We would like to thank E.J. Kolmes, M.E. Mlodik, and T. Rubin for helpful discussions.
This work was supported by DOE grant DE-SC0016072, DOE-NNSA grants DE-NA0003871,  DE-SC0021248, and 83228-10966 [Prime No. DOE (NNSA) DE-NA0003764], and NSF grant PHY-1506122.
One author (IEO) also acknowledges the support of the DOE Computational Science Graduate Fellowship (DOE Grant No. DE-FG02–97ER25308).

\section*{Data Availability}

Data sharing is not applicable to this article as no new data were created or analyzed in this study.

\appendix

\section{Collisions and Charge Transport in a Magnetized Plasma with $\ve{E} \times \ve{B}$ Flow Shear} \label{app:viscosity}

In this section, which draws from Chapter~3 of Ref.~\cite{Ochs2022Thesis}, we consider the charge transport that occurs when two particles collide in a magnetized plasma in the presence of an inhomogeneous electric field (and thus $\ve{E} \times \ve{B}$ flow shear).
The analysis in this section bears some similarity to that of Kaufman \cite{kaufman1960plasma}, in that it considers full orbits, but is also different in that it leverages Hamiltonian constants of motion rather than solving dynamical equations for large distributions of particles.
The development here thus better emphasizes the relationship between momentum conservation and charge transport in the single-particle picture.

To that end, consider a particle in a uniform magnetic field along $z$.
We want to examine the dynamics in the presence of a uniform magnetic field $\ve{B} \parallel \hat{z}$, and an electric field $\ve{E}(x) \parallel \hat{x}$ that varies in strength along $x$.

To simplify the algebra, we work in the frame that moves at the local $y$-directed (sub-relativistic) $\ve{E} \times \ve{B}$ velocity at $x = 0$.
The boost of the fields eliminates the electric field at $x=0$, without significantly changing the magnetic field.
Thus, Taylor expanding the electric field near $x$, we will only have electric field terms proportional to $x$ and $x^2$, and thus only have electric potential terms proportional to $x^2$ and $x^3$, and we have the vector and scalar potentials:
\begin{align}
	A_y &= B x; \quad \phi = - \phi_2 \lp \frac{x}{d} \rp^2 - \phi_3 \lp \frac{x}{d} \rp^3.
\end{align}
This leads to a Hamiltonian and canonical momentum given by:
\begin{align}
	H = \frac{1}{2} m v_x^2 + \frac{1}{2} m v_y^2 + q \phi; \quad p_y = m v_y + \frac{q}{c} A_y. \label{eq:viscHamMomentum}
\end{align}
Both of these are invariants of the orbit.
%

To get the gyrocenter position, we can combine these equations to eliminate $v_y$.
We can then solve for the upper and lower bounds of the orbit by setting $v_x = 0$.
The average of these bounds gives the gyrocenter position.

This process will look cleaner if we nondimensionalize.
Define:
\begin{alignat}{3}
	x &= d \bx; \quad & H &= \frac{m}{2} \Omega^2 d^2 \bH; \\ 
	p_y &= m \Omega d \bp; \quad &\phi_a &= \frac{1}{q} \frac{m}{2} \Omega^2 d^2 \bphi_a. \label{eq:hamiltonianViscosityDimensions}
\end{alignat}
Combining Eqs.~(\ref{eq:viscHamMomentum}) to eliminate $v_y$, and setting $v_x = 0$, we find the equation which determines the orbit limits:
\begin{align}
	\bH &= (\bx-\bp)^2 - \bx^2 \bphi_2 - \bx^3 \bphi_3. \label{eq:viscHamNondim}
\end{align}

Rather than list the whole cubic solution, we will solve this in orders in $\bphi_a \ll 1$.
To zeroth order, we have the orbit limits:
\begin{align}
	\bx_{\pm}^{(0)} = \bp \pm \sqrt{\bH}.
\end{align}
This corresponds to a gyrocenter position:
\begin{align}
	\bx_\text{gc}^{(0)} = \bp. \label{eq:singleParticleZerothOrder}
\end{align}

Eq.~(\ref{eq:singleParticleZerothOrder}) shows that, to lowest order, there is a linear relation between a particle's gyrocenter position and canonical momentum.
Thus, in the absence of spatially-varying electric field, when two particles with equal charge and mass exchange momentum, they will move an equal and opposite amount, and no net charge will be moved, as we found in Section~\ref{sec:momentumMagnetized}.

To explore what happens when the electric fields are added, we plug Eq.~(\ref{eq:singleParticleZerothOrder}) back into our Hamiltonian Eq.~(\ref{eq:viscHamNondim}), now looking at terms to first order.
Carrying out the same process of averaging to find the gyrocenter position, we find:
\begin{align}
	\bx_\text{gc}^{(1)} = \bp \bphi_2 + \frac{1}{2} \lp \bH + 3 \bp^2 \rp \bphi_3.
\end{align}
Now, we see, our former linear relationship between $\bx_{gc}$ and $\bp$ has become nonlinear, thanks to the $\phi_3$ term.

To see how this modification affects charge transport, consider a collision that averages momentum between two particles $a$ and $b$ with the same $\bH$, and with $\bp_a = -\bp_b = \bp_0$.
After the collision, both particles end up with $\bp_f = 0$, and retain $\bH_f = \bH_0$ (this corresponds to a $90^\circ$ collision).
The initial average gyrocenter position of the two particles is:
\begin{align}
	\llangle \bx_\text{gc,0} \rrangle_{ab} &= \frac{1}{2} \lp 3 \bp_0^2 + \bH \rp \bphi_3.
\end{align}
After the collision, the final gyrocenter average is:
\begin{align}
	\llangle \bx_\text{gc,f} \rrangle_{ab} &= \frac{1}{2} \bH \bphi_3.
\end{align}
Thus:
\begin{align}
	\Delta \llangle \bx_\text{gc} \rrangle &= -\frac{3}{2} \bp^2 \bphi_3. \label{eq:HamViscGyroShift}
\end{align}

One naturally wonders whether this charge transport corresponds to the perpendicular Braginskii viscosity\cite{Braginskii1965}, since it creates a current that might flatten out the $\ve{E} \times \ve{B}$ flow\cite{kolmes2019radial}.
To find out, we can estimate the current due to this process in a plasma.
Thus, take:
\begin{align}
	j_x \sim n q \Delta x_\text{gc} \nu,
\end{align}
where $n$ is the density, $q$ is the charge, and $\nu$ is the collision rate.
To estimate $\Delta x_\text{gc}$, we need an estimate for $\bp_0$, the difference in momentum of the colliding particles.
Luckily, we can see from our zeroth order solution that the nondimensional gyroradius is given simply by $\bar{\rho} = \sqrt{\bH}$. 
Thus, collisions between particles with a separation of a gyroradius correspond to $\bp_0 = \sqrt{\bH}/2$.
The typical particle energy will be $H \sim T$.
Finally, we note that $\phi_3$ is related to the electric field by:
\begin{align}
	\phi_3 = \frac{d^3}{6} \pa{^2E}{x^2}.
\end{align}
Putting this all together, using Eqs.~(\ref{eq:HamViscGyroShift}) and (\ref{eq:hamiltonianViscosityDimensions}), and taking our species to be ions, we have:
\begin{align}
	j_x &\sim \frac{c}{B} \pa{}{x} \left[\lp \frac{n_i T_i \nu_{ii}}{4 \Omega_i^2} \rp \pa{v_{E\times B}}{x} \right] \approx -\frac{c}{B} (\nabla \cdot \pi_i)_y,
\end{align}
where we have recognized the approximate form and scaling of the Braginskii perpendicular viscosity $\eta_1 \approx n_i T_i \nu_{ii} / 4 \Omega_i^2$, and the disagreement in the $\mathcal{O}(1)$ factors is due to the detailed kinetics of the collision operator.

Thus, charge transport is possible when there are inhomogeneous fields; however, it occurs at a rate ordered down by $(\rho/L)^2$ compared to the charge transport of each individual particle involved in the collisions.
Throughout most of this paper, we are concerned with whether charge transport occurs at 0th order in $\rho / L$, rather than 2nd order, and neglect these finite-Larmor-radius effects.

\section{Ponderomotive Forces: Comparing Fluid and Oscillation Center Theory} \label{app:ponderomotive}

Here, we briefly go through a calculation of the ponderomotive force on a plasma in both the fluid and quiver Lagrangian theories.
For a longer discussion with more intermediate steps, see Ref.~\cite{Ochs2022Thesis}, Chapter 7.

We start with the fluid quasilinear theory.
By combining the Reynolds stress and electromagnetic forces on the plasma, as done for the lower hybrid wave in Ref.~\cite{Ochs2021WaveDriven}, we can find the force on the fluid element from and electrostatic wave in an unmagnetized plasma, yielding:
\begin{align}
	F_{\text{fluid},s}^i &= 2 W_{EM}  \left[k_{r}^i  D_{is} + k_{r}^i \omega_i \pa{D_{rs}}{\omega_r} - \kappa^i D_{rs} \right],
\end{align}
where $\ve{\kappa}$ is the imaginary component of $\ve{k}$.
Here, the first term is the resonant force, the second term is the nonresonant recoil force for time-growing waves, and the third term is the gradient ponderomotive force.
In steady state, a pressure gradient must form to balance the ponderomotive force, canceling the last term.

Now that we have covered the fluid theory, we will proceed to a derivation of the force from an alternative perspective, using the quiver-Lagrangian theory \cite{Dodin2014Variational} of the oscillation center (OC).
The starting point for this theory is the Langrangian for a particle in an electric field:
\begin{align}
	\mathcal{L} = \frac{1}{2} m v^2 - q \phi(\ve{x},t).
\end{align}
Here, $\phi(x,t)$ is the space- and time-dependent potential associated with the electrostatic wave.

The first key step is to split the particle coordinate into two pieces: a small part $\ve{\tx}$ associated with the oscillation that changes quickly and averages to 0 over a wave cycle, and a large part $\ve{X}$ that corresponds to the center of the oscillation.
If we examine the Euler-Lagrange equations obtained by varying $\ve{\tx}$, and assume $d V^i / dt$ is $\mathcal{O} (E^2)$, then we find to first order in $E$:
\begin{align}
	m \frac{d\tv^i}{dt}  &=  q E^i(\ve{X}). \label{eq:eomvt}
\end{align}
We then assume the wave is locally monochromatic, so that:
\begin{align}
	E^i(\ve{X}) &= k^i \phi_a (\ve{X}) \sin(k_j x^j - \omega t + \theta),
\end{align}
Since $dV^i / dt$ is $\mathcal{O}(E^2)$, we have to first order $\ve{X} = \ve{X}_0 + \ve{V} t$. 
Plugging this in to Eq.~(\ref{eq:eomvt}) and solving yields:
\begin{align}
	\tv^i &= -\frac{q k^i \phi_a(\ve{X})}{m} \frac{1}{(k_j V^j - \omega)} \cos((k_j V^j - \omega) t + \theta') \label{eq:vQuiver}\\
	\tx^i &= -\frac{q k^i \phi_a(\ve{X})}{m} \frac{1}{(k_j V^j - \omega)^2} \sin((k_j V^j - \omega) t + \theta') \label{eq:xQuiver}.
\end{align}

Now we are ready to calculate the evolution of $(\ve{X},\ve{V})$.
Because we are only interested in changes on long timescales, we will average the full Lagrangian over an oscillation period.
This gives:
\begin{align}
	\bar{\mathcal{L}} \approx \frac{1}{2} m  V^2  + \frac{1}{2} m \llangle \tv^2 \rrangle  + q \llangle \ve{\tx} \cdot \ve{E} (\ve{X},t) \rrangle. \label{eq:ocLagrangianMiddle}
\end{align}
Plugging in for $\tx$ and $\tv$ and averaging, we find the oscillation center Lagrangian:
\begin{align}
	\bar{\mathcal{L}} \approx \frac{1}{2} m  V^2  - \Phi(\ve{X},\ve{V}), \label{eq:ocLagrangianFinal}
\end{align}
where we have defined the ponderomotive potential:
\begin{align}
	\Phi(\ve{X},\ve{V}) &= \frac{1}{4} \frac{q^2}{m} \frac{k^2 |\phi_a(\ve{X})|^2}{(\ve{k} \cdot \ve{V} - \omega)^2} \label{eq:pondPotential}
\end{align}

From this Lagrangian, the ponderomotive force is given by the Euler-Lagrange equations, which give to $\mathcal{O}(E^2)$:
\begin{align}
	m \frac{d V^i}{dt} = \frac{\partial^2\Phi}{\partial V_i \partial t} + \frac{\partial^2 \Phi}{\partial V_i \partial X_j} V_j - \pa{\Phi}{X_i}. \label{eq:pondForceSingleParticle}
\end{align}

\subsection{Integration of the Total Ponderomotive Force}

Now we seek to use Eq.~(\ref{eq:pondForceSingleParticle}) to derive the force on a volume of plasma.
To do this, we integrate the single-particle force over the distribution of oscillation centers $F$.
(Here, the variable $F$ plays double-duty as both a distribution and a force density. It should be clear from context which is which, since the force density will always be bolded or have a coordinate superscript.)
Taking $F$ to be normalized to 1, we have:
\begin{align}
	F_{s,\text{p}}^i &= n_{s0} m_s \pv \int d\ve{V} F_s \frac{dV^i}{dt}
\end{align}
Making use of Eqs.~(\ref{eq:pondForceSingleParticle}-\ref{eq:pondForceSingleParticle}), the definition of $W_{EM}$ in Eq.~(\ref{eq:wpEM}), and the fact that $\paf{W_{EM}}{x^j} = -2\kappa_j W_{EM}$ and $\paf{W_{EM}}{x^j} = 2\omega_i W_{EM}$, we can eventually write this as:
\begin{align}
	F_{s,\text{p}}^i &= 2 W_{EM}  \lp k_i \lp \omega_i \pa{}{\omega_r} + \kappa_j \pa{}{k_j} \rp - \kappa_j \delta_{ij}\rp \notag\\
	& \hspace{0.5in} \times \pv\int d\ve{V} \lp -\frac{\omega_{ps}^2 F_s}{(\omega-\ve{k} \cdot \ve{V})^2} \rp .
\end{align}
We can recognize the integral on the second line from electrostatic Vlasov-Poisson theory \cite{kralltrivelpiece,stix1992waves} as the real part of the dispersion relation contribution $D_{rs}$ from species $s$. 
Thus:
\begin{align}
	F_{s,\text{p}}^i &= 2 W_{EM} \left[ k^i \lp \omega_i \pa{D_{rs}}{\omega_r} + \kappa_j \pa{D_{rs}}{k_j} \rp - \kappa^i D_{rs} \right]. \label{eq:fPondUnmag} 
\end{align}
Again, note also that this equation has only been shown for the special case of electrostatic waves in an unmagnetized plasma, and is not a general relation for all electrostatic waves.

\subsection{Reynolds and Polarization Stress}

As shown by Gao \cite{Gao2006,Gao2007}, for a cold fluid, the fluid and oscillation center forces in general differ
by the Reynolds stress and the polarization stress.
The discrapancy arises because the force we have calculated thus far is the integrated force over a set of oscillation centers.
If we consider the difference in the force applied to the particle distribution in a volume vs. the oscillation center distribution in a volume, there are two sources of discrepancy.

First, the oscillation center does not oscillate in the wave, while the particle does. 
Thus there will be a Reynolds stress (and resulting macroscopic force) associated with the particle oscillation that is absent for the oscillation center.
For a given oscillation center distribution $F(\ve{X},\ve{V})$, this stress is given to $\mathcal{O}(E^2)$ by:
\begin{align}
	\Pi_\text{Rey}^{ij} &= m n_0 \int d\ve{V} F(\ve{X},\ve{V}) \llangle \tv^i \tv^j \rrangle.
\end{align}

Second, just because the oscillation center is within a certain volume does not mean that its associated particle is within the same volume (or vice versa), since the particle is displaced from the oscillation center by $\ve{\tx}$.
Since the displacement $\ve{\tx}$ is correlated with the electric field $\ve{E}$, the forces from lost and added particles will sum, and lead to a surface stress.

To get an explicit expression for the polarization stress, consider a cube of plasma, and the boundaries along the $\ve{x}^j$ direction.
At any given position on the interface, there will be an excess surface charge density (relative to that from the oscillation centers) $\sigma_{xs} = s \int d\ve{V} F(\ve{X},\ve{V}) q n_0 \tx^j$ of charge, due to charge entering or leaving the interface.
Here, $s$ is a sign that depends on the interface, with $s = 1$ at the low-$x^j$ boundary, and $s=-1$ at the high-$x^j$ boundary.
This charge combines with the electric field to produce a force on the fluid element, which we obtain by integrating over the coordinates that lie in the constant-$x^j$ surface, and averaging over the wave cycle:
\begin{align}
	\int d\ve{X} F_\text{pol}^i &= -\int dX^{m\neq j} \llangle E^i \sigma_{xs} \rrangle \\
	&= -\int dX^{m\neq j}\int d\ve{V} F(\ve{X},\ve{V}) q n_0 \llangle \tx^j E^i \rrangle \biggl|^{x^j_1}_{x^j_0}.
\end{align}
Applying Stokes' theorem along the $x^j$ coordinate allows us to express this as the divergence of a polarization stress tensor:
\begin{align}
	\int d\ve{X} F_\text{pol}^i &=  \int d\ve{X} \lp -\pa{}{x^j} \Pi^{ij}_\text{pol} \rp\\
	\Pi^{ij}_\text{pol} &\equiv q n_0 \int d\ve{V} F(\ve{X},\ve{V})  \llangle \tx^j E^i \rrangle.
\end{align}  
Using our solutions for $\tx$ and $\tv$, it is possible to show:
\begin{align}
	\Pi_\text{Rey}^{ij} = -\Pi_\text{pol}^{ij} = -2W_{EM} \frac{k^i k^j}{k^2} D_{rs}.
\end{align}
Thus, the two stresses in this case cancel out, so that the OC and fluid forces should agree.

Now we are in a position to get the total force on the plasma in OC theroy, by summing Eq.~(\ref{eq:fPondUnmag}) with the force on resonant particles from Eq.~(\ref{eq:QLforce1D}), $F_{s,\text{res}}^i = 2W_{EM} k^i D_{is}$.
Summing up the forces on all species, and making use of the imaginary and real parts of the dispersion relation (Eqs.~(\ref{eq:realDisp}-\ref{eq:imagDisp})), we have:
\begin{align}
	\sum_s \lp F_{s,\text{p}}^i + F_{s,\text{res}}^i \rp  =  2 W_{EM}  \kappa^i.
\end{align}
Importantly, we see that this force only points along the direction of wave decay, and thus (in contrast to the fluid force) cannot give a force along $\theta$ in a cylindrically symmetric plasma.

However, in performing this analysis, we have left out one important term, absent in the cold fluid theory: the stress of the oscillation center distribution itself.
As it turns out, in steady state, the self-consistent oscillation center distribution is anisotropic, leading to a stress that allows for agreement with the fluid theory.

\subsection{Oscillation Center Stress in Steady State} \label{sec:ocStressLagrangian}

To get the stress in steady state, we need to calculate the OC distribution.
To do this, we take the distribution function far from the wave to be Maxwellian, since this is the region of unperturbed plasma.
We can then solve for the distribution in the wave region by using a combination of invariants of the motion, and incompressibility of the phase space.
(For a calculation with more intermediate steps, see Ref.~\cite{Ochs2022Thesis}, Chapter 7).

Our OC Lagrangian (Eq.~(\ref{eq:ocLagrangianFinal})) can be written:
\begin{align}
	\mathcal{L} &= \frac{1}{2} m V^2 - \frac{C}{(1-k^i V_i / \omega)^2}; \quad C \equiv \frac{\omega_p^2}{\omega^2} \frac{W_{EM}}{n_0}.
\end{align}
For simplicity, we will only consider waves for which $\omega / k \gg V$, (a) so that we don't have to worry about particles encountering a Landau resonance, and (b) so that we can Taylor expand:
\begin{align}
	\mathcal{L} &\approx \frac{1}{2} m V^2 - C \lp 1 + 2 \frac{k_i V^i}{\omega} + 3 \frac{(k_j V^j)^2}{\omega^2}\rp.
\end{align}

This Lagrangian allows us to find the canonical momentum:
\begin{align}
	P_i &\equiv \pa{\mathcal{L}}{V^i} = m V_i - 2 C \frac{k_i}{\omega} - 6 C \frac{k_i k_j V^j}{\omega^2}
\end{align}

The Hamiltonian is then given by:
\begin{align}
	\mathcal{H} &= P_i V^i - \mathcal{L} = \frac{1}{2} m V^2 + C\lp 1 - 3 \frac{(k_i V^i)^2}{\omega^2} \rp.  
\end{align}

Using the definition of $P_i$, this can be rewritten in terms of $P_i$ to $\mathcal{O} (C)$ as:
\begin{align}
	\mathcal{H} &= \frac{P^2}{2m} + C + 2 C \frac{k_i P^i}{m \omega} + 3 C \frac{(k_j P^j)^2}{m^2 \omega^2}.
\end{align}

Now, we take $C$ to be independent of $Y$.
This makes $P_Y$ an invariant of the dynamics.
Furthermore, because we are examining the problem in steady state, $\mathcal{H}$ is also an invariant of the dynamics.
The two constraint equations for these two invariants allow us to determine $P_{X0}$ and $P_{Y0}$ in terms of $P_{X}$ and $P_{Y}$, where $P_{X0}$ and $P_{Y0}$ correspond to an unperturbed region far from the wave.
The solutions are:
\begin{align}
	P_{Y0} &= P_Y\\
	P_{X0} &\approx P_X + \frac{mC}{P^X} + 2 C \frac{k_i P^i}{\omega P^X} + 3 C \frac{(k_j P^j)^2}{m \omega^2 P^X} .
\end{align}

Now we are in a position to calculate the distribution function in the wave region.
Far from the wave, we will take the distribution to be Maxwellian, so that:
\begin{align}
	F_{P0} (\ve{P}_0) &= \frac{1}{2 \pi m^2 v_{th}^2} e^{-(P_{X0}^2 + P_{Y0}^2)/2m^2 v_{th}^2}
\end{align}

Then, in the wave region, phase space incompressibility gives us:
\begin{align}
	F_{P}(\ve{P}) &= F_{P0} (\ve{P}_0(\ve{P})) \approx F_{P0} (\ve{P}) + \pa{F_{P0}}{P_i} (P_{i0} - P_i).
\end{align}
For our problem, we have $P_{Y0} = P_Y$, so the second term comes purely from variation in $P_X$.
For a Maxwellian, our derivative is given by:
\begin{align}
	\pa{F_{P0}}{P_i} &= -\frac{P^i}{m^2 v_{th}^2} F_{P0}.
\end{align}
Thus,
\begin{align}
	F_P(\ve{P}) &\approx F_{P0}(\ve{P}) \left[1 - \frac{C}{m v_{th}^2} \lp 1 + 2 \frac{k_i P^i}{m \omega} + 3 \frac{(k_j P^j)^2}{m^2 \omega^2} \rp \right].
\end{align}

To take the relevant moments, it is convenient to transform this distribution from momentum space $\ve{P}$ to velocity space $\ve{V}$.
This transformation is given by:
\begin{align}
	F_V(\ve{V}) &= \frac{\sqrt{g^V}}{\sqrt{g^P}} F_P(\ve{P}(\ve{V})).
\end{align}
where $g^{P}_{ij} = \delta_{ij}$, and:
\begin{align}
	g^V_{ij} &= \pa{P^m}{V^i}\pa{P^n}{V^j} g^P_{mn}.
\end{align}
After taking the relevant derivatives and performing some algebra, we arrive at (to $\mathcal{O}(C)$)
\begin{align}
	F_V(\ve{V}) &= \frac{e^{-V^2 / 2 v_{th}^2}}{2\pi  v_{th}^2}   \biggl[1 +  \frac{W_{EM}}{p_{th0}} \frac{\omega_p^2}{\omega^2} \notag\\
	& \hspace{0.5in} \times \lp -1 - 6 \frac{k^2 v_{th}^2}{\omega^2} + 3 \frac{(k_j V^j)^2}{\omega^2} \rp\biggr], \label{eq:ocFvFinal}
\end{align}
where $p_{th0} \equiv n_0 m v_{th}^2 = n_0 T$.

Thus, we see that the effect of the field is to overall reduce the value of $F_V(\ve{V})$ in regions with a wave.
This repulsion of particles from the wave region due to the ponderomotive potential is unsurprising.
However, we also see that the wave introduces anisotropy to the velocity-space oscillation center distribution, which stretches out somewhat along $\ve{k}$ thanks to the last term (Fig.~\ref{fig:ocDistribution}).

Now we can calculate our moments.
With a bit of algebra, we find:
\begin{align}
	n &=  n_0 \left(1 + \frac{W_{EM}}{p_{th0}} D_{rs} \right) \label{eq:ocDensity}\\
	n u^i &= n_0 \int d\ve{V} V^i F_V(\ve{V}) = 0\\
	\Pi_{oc}^{ij} &= \lp p_{th0} + W_{EM} D_{rs} \rp \delta^{ij} - W_{EM} k^i \pa{D_{rs}}{k_j}.
\end{align}

This stress leads to a force:
\begin{align}
	F_{s,\Pi}^i &= -\pa{}{x^j} \Pi^{ij}_{oc}  =  2 W_{EM} \lp \kappa^i D_{rs} - k^i \kappa_j  \pa{D_{rs}}{k_j} \rp \label{eq:ocStressForce}
\end{align}
Comparing to Eq.~(\ref{eq:fPondUnmag}), we see that this force exactly cancels the ponderomotive force on nonresonant particles in steady state.
Thus, along the symmetry direciton $y$, only the (uncompensated) force on the resonant particles remains, in agreement with the fluid theory of the main text.


%


\clearpage

\end{document}